\documentclass[prd,twocolumn,superscriptaddress,floatfix,showpacs,nofootinbib]{revtex4-2}
\usepackage{amsfonts}
\usepackage{amsmath}
\usepackage{amssymb}
\usepackage{xcolor}
\usepackage{graphicx}
\usepackage[normalem]{ulem}
\font\mybb=msbm10 at 10pt
\def\bb#1{\hbox{\mybb#1}}
\def\be{\begin{equation}}
\def\ee{\end{equation}}
\def\bea{\begin{eqnarray}}
\def\eea{\end{eqnarray}}
\renewcommand{\theequation}{\arabic{section}.\arabic{equation}}
\begin{document}
\begin{flushright}
V2: 2025, October 15. JHEP 12(2025)003
\end{flushright}
\bigskip
\title{Quantum state spectrum and field theory of D$0$-brane
}

\author{Igor Bandos}
\email{igor.bandos@ehu.eus}
\affiliation{Department of Physics and EHU Quantum Center, University of the Basque Country UPV/EHU,
P.O. Box 644, 48080 Bilbao, Spain,}
\affiliation{IKERBASQUE, Basque Foundation for Science,
48011, Bilbao, Spain, }

\author{Unai D.M. Sarraga}
\email{unai.demiguel@ehu.eus}
\affiliation{Department of Physics and EHU Quantum Center, University of the Basque Country UPV/EHU,
P.O. Box 644, 48080 Bilbao, Spain,}
\affiliation{Institute for Theoretical Physics UAM/CSIC, C/ Nicolás Cabrera 13-15,
Campus de Cantoblanco, Universidad Autonoma de
Madrid,
28049 Madrid Spain}

\author{Mirian Tsulaia}
\email{mirian.tsulaia@oist.jp}
\affiliation{Okinawa Institute of Science and Technology,
1919-1 Tancha, Onna-son, Okinawa 904-0495, Japan}

\bigskip

\begin{abstract}
We quantize D$0$-brane, the simplest representative of the family of supersymmetric Dirichlet $p$-branes, in its spinor moving frame formulation and analyze its quantum state spectrum. Besides being a preparatory stage for quantization of the complete supersymmetric multiple D$0$-brane (mD$0$) model, which is presently known only in the frame of spinor moving frame formalism, this is interesting as its own: despite the covariant quantization of $\text{D}=10$ D$0$-brane was discussed before, its quantum state spectrum was not analyzed and its field theory, describing that, was not studied.

We show that the quantum state vector of the D$0$-brane is described by an on-shell superfield that collects the fields of the massive counterpart of the linearized type IIA supergravity supermultiplet which can be obtained as massive Kaluza-Klein mode of the dimensional reduction of the eleven dimensional supergravity down to ten dimensions. In conclusion, we briefly discuss a spinor helicity formalism for D$0$-brane which can be used as one of the basic elements for developing type IIA string theory amplitude calculus based on multiparticle generalizations of the on-shell superfields.
\end{abstract}

\maketitle
\begin{widetext}
\tableofcontents

\setcounter{equation}0

\section{Introduction}
Dirichlet $p$-branes, also called D$p$-branes, were discovered (theoretically) in  \cite{Sagnotti:1987tw,Dai:1989ua,Horava:1989ga} being hitherto ignored  types of boundary conditions for an open fundamental string \cite{Goddard:1973qh}. String ending on D$p$-brane has Dirichlet boundary conditions in $\text{D}-p-1$ directions orthogonal to a $p+1$ dimensional plane  ($\text{D}=10$ for critical superstrings) and Neumann boundary conditions in $p+1$ directions tangential to this plane. The limited case of spacetime filling D$9$-brane, with Neumann boundary conditions in all directions, corresponds to an open string and was known much earlier \cite{Goddard:1973qh,Brink:1976sc}.

Already in the first papers \cite{Sagnotti:1987tw,Dai:1989ua,Horava:1989ga} it was appreciated that, since the String Theory is a theory of gravity, it cannot contain rigid surfaces, therefore the D$p$-branes should actually be dynamical objects. In supersymmetric string theory these should be supersymmetric extended objects, like fundamental the superstring itself (hence the longer name ``super-D$p$-branes", which was often used for D$p$-branes).

In the seminal paper \cite{Polchinski:1995mt} J. Polchinski identified super-D$p$-branes with supersymmetric $p$-brane solutions of $\text{D}=10$ ${\cal N}=2$, type IIA or type IIB, supergravity which had  seemed  mysterious \cite{Duff:1992hu} before. Their  worldvolume actions were found a bit later in \cite{Cederwall:1996pv,Aganagic:1996pe,Cederwall:1996ri,Aganagic:1996nn,Bergshoeff:1996tu,Bandos:1997rq}\footnote{The action for super-D$0$-brane, elaborated in \cite{Bergshoeff:1996tu} is a generalization to the type IIA superspace of $\text{D}=4$ ${\cal N}=2$ superparticle action by J. A. de Azc\'arraga and J. Lukierski \cite{deAzcarraga:1982dhu}.}; these actions possess both spacetime supersymmetry and local fermionic $\kappa$-symmetry which indicates that the ground state of the dynamical system preserves a part of supersymmetry\footnote{Notice that the equations of motion of D$p$-branes were found a bit earlier by Howe and Sezgin \cite{Howe:1996mx} in the frame of superembedding approach \cite{Bandos:1995zw,Sorokin:1999jx,Bandos:2023web}. In this approach the local fermionic $\kappa$-symmetry \cite{deAzcarraga:1982dhu,Siegel:1983hh} is represented by worldvolume superspace supersymmetry, as noticed for the first time in superparticle models in \cite{Sorokin:1988jor}.}.

Notice that in 10D supergravity theories, being various low energy limits of String Theory, D$p$-brane solutions  carry electric or/and magnetic charges of the Ramond-Ramond (RR) fields of 10D type II supergravity supermultiplets \cite{Polchinski:1995mt}. This is a result  of the solitonic nature of these objects in the standard approach to the theory of fundamental strings. However, in the frame of the Brane Democracy concept \cite{Townsend:1995gp}\footnote{This generalizes the Mantonen-Olive duality conjecture \cite{Montonen:1977sn}.  }, any of the $p$-branes, in particular of D$p$-branes, can be considered as fundamental objects of String/M-theory and the others should appear as solitons in this $p$-brane based approach. In this respect, D$0$-branes were of  special interest because their existence suggested that we could reformulate String Theory as a theory of particles.

An important property of D$p$-branes is that they carry a gauge field on their worldvolume. This makes them an essential component of (semi)realistic models of our Universe
derived from String Theory (see e.g.  \cite{Cvetic:2011vz,Ibanez:2012zz}).
However, to realize the Standard Model on the D$p$-brane worldvolume, the gauge fields should be non-Abelian. This implies that a phenomenologically interesting  Brane World model can be constructed rather on a (center of mass) worldvolume of  system of several ($N$) nearly coincident D$p$-branes.  We can call this system multiple D$p$-brane or mD$p$ system.

Such a mD$p$ system includes, in addition to $N$ nearly coincident D$p$-branes, also $N^2$ strings ending on the same or on a different brane. At low energies this system can be described by $N^2$ gauge fields on the $\text{d}=(p+1)$ dimensional center of mass worldvolume. At the first glance, it appears we have $N^2$ Abelian gauge fields.
However,
in \cite{Witten:1995im} Witten argued that  the enhancement of gauge symmetry from   $\text{U}(1)^{N^2}$  to $\text{U}(N)$ occurs and the system  is described in terms of fields of  maximally supersymmetric  $\text{d}=(p+1)$ dimensional {\it non-Abelian gauge theory}.

At  low energy the system can be described   by $\text{U}(N)$ super-Yang-Mills (SYM) action
\cite{Witten:1995im}. However, when the field strength is not so small (but not quickly varying so that derivative of field strengths can be neglected) one can argue that some higher-order terms should be present in the action. Indeed, in the particular case of $N=1$, corresponding to a single D$p$-brane, the complete action is  given  by the sum of the Dirac-Born-Infeld and the Wess-Zumino terms \cite{Cederwall:1996pv,Aganagic:1996pe,Cederwall:1996ri,Aganagic:1996nn,Bergshoeff:1996tu,Bandos:1997rq} and $\text{U}(1)$ the SYM action can be obtained as weak field limit of the gauge fixed form of this action.

To summarize, the quest for the complete multiple D$p$-brane  action will soon reach the age of 30 years. Although many interesting results have been obtained over these years
\cite{Tseytlin:1997csa,Emparan:1997rt,Taylor:1999gq,Taylor:1999pr,Myers:1999ps,Millar:2000ib,Bergshoeff:2000ik,
Bergshoeff:2001dc,Sorokin:2001av,Janssen:2002vb,Drummond:2002kg,Panda:2003dj,Janssen:2002cf,Janssen:2003ri,
Lozano:2005kf,Howe:2005jz,Howe:2006rv,Howe:2007eb,Bandos:2009yp,Bandos:2009gk,Bandos:2010hc,McGuirk:2012sb,
Bandos:2012jz,Bandos:2013uoa,Choi:2017kxf,Choi:2018fqw,Bandos:2018ntt,Brennan:2019azg,Bandos:2021vrq,
Bandos:2022uoz,Bandos:2022dpx}, (see in particular \cite{Bandos:2022uoz,Bandos:2022dpx} for a comprehensive discussion of this subject),  the problem has not been solved yet, at least in its complete form.

Below we briefly describe  the results of \cite{Bandos:2022uoz,Bandos:2022dpx}  as they provide one of the motivations for our present study. In  \cite{Bandos:2022uoz} the supersymmetric and $\kappa$-symmetric action possessing the properties expected for the complete action of  multiple D$0$-brane (mD$0$) system  has been constructed. Its properties and the $11$D origin were studied in   \cite{Bandos:2022dpx}. The appropriate quantization of this dynamical system will supposedly give the mD$0$ field theory. This is a field theory in superspace with additional non-commutative coordinates, the study of which might provide an important insight into String/M-theory.

The problem of quantization of mD$0$ action of \cite{Bandos:2022uoz,Bandos:2022dpx} is nontrivial in many respects and one of the motivations for the present study is to create the basis for such quantization.

In this paper we will quantize the single D$0$-brane model in its spinor moving frame formulation \cite{Bandos:2000tg}.  Besides being the first step towards  the mD$0$ supersymmetric field theory, such a study is interesting on its own. Although the quantization of D$p$-branes and, particularly of D$0$-branes, were discussed in the literature  \cite{Kallosh:1997nr}, to the best of our knowledge  the quantum state spectrum of D$0$-brane has not been analyzed, and the corresponding field theory has been studied before. Furthermore, the quantization in spinor moving frame formulation allows us to elaborate the spinor helicity formalism for type IIA scattering amplitudes which involve D$0$-branes. In concluding Sec.~\ref{Conclusion}  we briefly discuss the spinor helicity formalism for D$0$-brane in its relation with spinor moving frame formulation.

\setcounter{equation}0

\section{D$0$-brane in spinor moving frame formulation. Classical mechanics and quantization}
\label{D0action=Sec}

\subsection{D$0$-brane action in spinor moving frame formalism. Central coordinate basis  }
\label{D0action}
Let us denote the bosonic and fermionic coordinates  of flat type IIA superspace by
\be\label{ZM} Z^M=( x^\mu, \theta^{\alpha 1}, \theta_{\alpha}^2) \; \qquad \mu=0,\ldots,9\; , \qquad \alpha =1,\ldots, 16 \;  \qquad  \ee
and define supersymmetry transformations
\begin{equation}\label{susy=IIA}
\delta x^\mu =i\theta^1\sigma^\mu \epsilon^1 +i\theta^2\tilde{\sigma}^\mu \epsilon^2 \; , \qquad \delta  \theta^{\alpha 1}= \epsilon^{\alpha 1}  \; , \qquad \delta  \theta^2_{\alpha}= \epsilon^2_{\alpha} \; , \qquad
\end{equation}
where
$\sigma^\mu_{\alpha\beta}=\sigma^\mu_{\beta\alpha}$ and  $\tilde{\sigma}{}^{\mu\alpha\beta}=\tilde{\sigma}{}^{\mu\beta\alpha}$ are 10D generalized Pauli matrices which obey
\be
\sigma^\mu\tilde{\sigma}{}^\nu+ \sigma^\nu\tilde{\sigma}{}^\mu = \eta^{\mu\nu} {\bb I}_{16\times 16}\; , \qquad \eta^{\mu\nu} ={\rm diag} (1,-1,\ldots, -1)\; .
\ee
The Volkov-Akulov 1-form invariant under type IIA supersymmetry \eqref{susy=IIA} reads
\begin{equation}
\Pi^\mu = \text{d}x^\mu -i\text{d}\theta^1\sigma^\mu \theta^1 -i\text{d}\theta^2\tilde{\sigma}^\mu \theta^2 \qquad.
\end{equation}
The action for a D$0$-brane  in the spinor moving frame formulation   \cite{Bandos:2000tg} can be obtained from the following first
order action involving an auxiliary unit timelike vector $u^0_\mu$
\begin{equation}\label{eq:LD0}
S^{\text{D}0} = \int\limits_{\mathcal{W}^1}{\cal L}_{1}^{\text{D}0}\equiv \int \text{d}\tau L^{\text{D}0} = m\int\limits_{\mathcal{W}^1}\Pi^\mu u^0_\mu -im \int\limits_{\mathcal{W}^1} \left(\text{d} \theta^{\alpha 1}\theta_{\alpha}^{2} - \theta^{\alpha 1}\text{d}\theta_{\alpha}^{2} \right)~, \qquad \qquad u^0_\mu u^{0\mu}=1\; .
\end{equation}
To this end  we consider this unit vector as a part of moving frame
\bea\label{harmU=m}
(u_\mu^{0}, u_\mu^{I})\in \text{SO}(1,9) \; , \qquad
 \eea
which is composed from spinor moving frame matrix
\be\label{v=inSpin}
v_\alpha{}^q \in \text{Spin}(1,9) \; ,  \qquad \alpha =1,\ldots, 16 \; ,  \qquad q =1,\ldots, 16
\ee
in the sense of the following constraints (see \cite{Bandos:2022uoz,Bandos:2022dpx} for more details)

\begin{eqnarray}\label{u0s=vv}
u_\mu^{0} \sigma^\mu_{\alpha\beta}=v_\alpha{}^q v_\beta{}^q \; ,  \qquad {}\qquad 
u_\mu^{I} \sigma^\mu_{\alpha\beta}=v_\alpha{}^q \gamma^I_{qp}v_\beta{}^p \; , \qquad  \nonumber \\ \nonumber \\ 
v_{\alpha}^q \tilde{\sigma}{}_{\mu}^{\alpha\beta}v_{\beta}^p= u_\mu^{0} \delta_{qp}+u_\mu^{I} \gamma^I_{qp}\; ,  \qquad I=1,\ldots,9\; . \qquad  \\ \nonumber 
\end{eqnarray}
Here   $\gamma^I_{qp}$ are 9d Dirac matrices which are symmetric, traceless and obey the Clifford algebra

\begin{eqnarray}\label{gIgJ=dIJ}
\gamma^I_{qp} =\gamma^I_{pq} \; ,  \qquad \gamma^I\gamma^J+\gamma^J\gamma^I=2\delta^{IJ} {\mathbb I}_{16\times 16} \; .  \qquad
\end{eqnarray}

These constraints, besides  implying \eqref{v=inSpin}, on the other side, result in that the composite vectors satisfy the equations

\begin{eqnarray}
\label{u0u0=1}
u^{\mu 0}u_\mu^{0}=1\; , \qquad u^{\mu 0} u_\mu^{I}=0\; , \qquad u^{\mu I} u_\mu^{J}=-\delta ^{IJ}\;
\\ \nonumber \\  \text{and}\qquad
\label{eu-u=e}
\epsilon^{\mu_1\ldots \mu_{10}}u_{\mu_1}^{0} u_{\mu_2}^{i_1}\ldots  u_{\mu_{10}}^{i_9} =
\epsilon^{i_1\ldots i_{9}}\;. \\ \nonumber
\end{eqnarray}
the set of which is equivalent to \eqref{harmU=m}.

The constraints \eqref{u0s=vv} imply $u_{\mu}^{{\rm 0}}= \frac 1 {16} v_q\sigma_\mu v_q$; substituting this in
\eqref{eq:LD0} makes manifest that the spinor moving frame action provides the massive D=10 generalization of the
Ferber-Schirafuji action \cite{Ferber:1977qx,Shirafuji:1983zd}  for D=4 massless superparticle. This latter provides a dynamical basis for twistor approach [71-74]. The literal generalization of \cite{Ferber:1977qx,Shirafuji:1983zd} for D=4 massless superparticle was done in \cite{Bette:2004ip,Fedoruk:2005ks} (a bit later than \cite{Bandos:2000tg}).

The integration in spinor moving frame action \eqref{eq:LD0} is over the worldline Here $\mathcal{W}^1$ i which can be defined as a line in type IIA parametrically using the coordinate  functions dependent on
of the proper time variable $\tau$
\bea \label{cW1}
 \mathcal{W}^1 \in \Sigma^{(10|32)} &: &\qquad Z^M=Z^M(\tau)=( x^\mu(\tau), \theta^{\alpha 1}(\tau), \theta_{\alpha}^2 (\tau))\;  . \qquad
\eea
This action  is invariant under the following  $\kappa$-symmetry transformations with local fermionic parameter  $\kappa^q(\tau)$ as

\begin{eqnarray}\label{eq:kappaD0=}
\delta_\kappa \theta^{\alpha 1}=   \kappa^q v_q{}^\alpha ~,~~~~~ \delta_\kappa \theta_{\alpha}^{2}= -  \kappa^q v_\alpha{}^q~,
\qquad  \delta_\kappa x^\mu =i\delta_\kappa\theta^1 \sigma^\mu \theta^1+i\delta_\kappa\theta^2 \tilde{\sigma}{}^\mu \theta^2~, \qquad
\nonumber \\ \nonumber \\
\delta_\kappa v_\alpha{}^q=0 \qquad \Longrightarrow \qquad \delta_\kappa u^I_\mu = \delta_\kappa u^0_\mu = 0~.
\end{eqnarray}
Here $ v_q{}^\alpha$ is inverse to spinor frame matrix $ v_\alpha{}^q$, 

\be\label{v-1=inSpin}
 v_\beta{}^p v_p{}^\alpha = \delta{}_\beta{}^\alpha
 \qquad \Longleftrightarrow \qquad v_p{}^\alpha v_\alpha{}^q = \delta_{pq} \qquad \Longrightarrow \qquad v_q{}^\alpha  \in \text{Spin} (1,9) \;  . \qquad
\ee

The local worldline supersymmetry leaving invariant the candidate
multiple D$0$-brane (mD$0$) action of \cite{Bandos:2022uoz,Bandos:2022dpx} is based on a generalization of the form of local fermionic $\kappa$-symmetry of a single D$0$-brane presented in \eqref{eq:kappaD0=}. Hence the quantization of single D$0$-brane in spinor moving frame formulation can be considered as a basis for quantization of such mD$0$ action. However, as we have already argued in the Introduction, there are a number of other reasons to perform such quantization, which which we briefly describe in the present paper.


It is useful to note that the D$0$-brane in its spinor moving frame formulation lives on the worldline in an enlarged superspace with coordinates
\be\label{LH=SSP}
\Sigma^{(10+45|16+16)}\; : \qquad {\cal Z}^{\cal M} =(Z^M , v_\alpha{}^q)= (x^\mu, \theta^{\alpha 1 },  \theta_{\alpha}^2, v_\alpha{}^q)
\ee
which is called the Lorentz harmonic superspace \cite{Bandos:1990ji}\footnote{In the present context the name ``light-cone" superspace used  in the seminal papers \cite{Sokatchev:1985tc,Sokatchev:1987nk} devoted to massless superparticle mechanics, cannot be applied.
Notice also that when writing the number of bosonic and fermionic degrees of freedom we use the notation for the superspace as $\Sigma^{(10+45|16+16)}$, to account for the fact that $v_\alpha^q$ obeys the constraints \eqref{v=inSpin} and hence provides generalized coordinates for the double covering $\text{Spin}(1,9)$ of the Lorentz group $\text{SO}(1,9)$.}.
The worldline of the superparticle is then defined parametrically in terms of the coordinate functions associated with this superspace, which we denote  by the same symbol as the coordinates
\be\label{WinLH=SSP}
{\cal W}^1 \in \Sigma^{(10+45|16+16)}: \qquad {\cal Z}^{\cal M}={\cal Z}^{\cal M}(\tau) = (x^\mu (\tau), \theta^{\alpha 1 } (\tau),  \theta_{\alpha}^2 (\tau),  v_\alpha{}^q(\tau))\; .
\ee
To be precise, we should call the coordinate $v_\alpha{}^q$ parameterizing $\text{Spin}(1,9)$ sector of the superspace \eqref{LH=SSP} {\it Lorentz harmonic} and reserve the name {\it spinor moving frame variables} for the constrained 1d field $ v_\alpha{}^q(\tau)$. However, we will not restrain ourselves to this rigorous terminology and often use these two names as synonyms.


The moving frame and spinor moving frame variables are highly constrained by Eqs.~\eqref{harmU=m} and \eqref{v=inSpin} which are equivalent, respectively, to Eqs.~\eqref{u0u0=1}, \eqref{eu-u=e} and Eqs.~\eqref{u0s=vv}. The simplest way to work with this is to apply the concept of admissible variations \cite{Bandos:1992ze,Bandos:1993yc} based on the use of Cartan forms
\bea\label{OmI:=}
 \Omega^{I}=u_\mu^{(0)} \text{d}u^{\mu I}=-\text{d}u_\mu^{(0)} u^{\mu I} \; , \qquad   \Omega^{IJ}= u_\mu^{I} \text{d}u^{\mu J} =-u_\mu^{J} \text{d}u^{\mu I}\; .
\eea
The second parts of the above equations  explicitly use the constraints \eqref{u0u0=1} on moving frame variables.
Referring to \cite{Bandos:2022uoz,Bandos:2022dpx,Bandos:2024hnj} (and refs. therein) for further discussion on the admissible variations and derivatives of the spinor frame and spinor moving frame variables, we will restrict ourselves by writing several equations which will be useful below.

The derivatives of the moving frame vectors and spinor moving frame variables can be expressed in terms of the Cartan forms by
\begin{eqnarray}\label{du0=}
\text{d}u_\mu^0 =u_\mu^I \Omega^I\; , \qquad \text{D}u^I_\mu := \text{d}u_\mu^I + u_\mu^J\Omega^{JI} =u_\mu^0 \Omega^I\; , \qquad
\\ \nonumber \\ \label{Dv=vOm}
\text{D}v_\alpha{}^q:= \text{d}v_\alpha{}^q+ \frac 1 4 \Omega^{IJ} v_\alpha{}^p\gamma^{IJ}_{pq}
= \frac 1 2 \gamma^I_{qp} v_\alpha{}^p\Omega^I \; . \qquad \\ \nonumber 
\end{eqnarray}
In these equations we have also introduced the $\text{SO}(9)$ covariant derivatives in which the Cartan form $\Omega^{JI}$ serves as a (composite) connection. The other Cartan form $\Omega^I$ is covariant with respect to $\text{SO}(9)$ gauge transformations.

The independent admissible variations of moving frame vectors and  spinor moving frame variables can be  parametrized  by the formal contractions of Cartan forms with the variational symbol,

\bea\label{ivOmI:=}
i_\delta \Omega^{I}=u_\mu^{(0)}\delta u^{\mu I}=-\delta u_\mu^{(0)}  u^{\mu I} \; , \qquad  i_\delta \Omega^{IJ}= u_\mu^{I} \delta u^{\mu J} =-u_\mu^{J} \delta u^{\mu I}\;
\eea
and  are thus given by formal contractions $ i_\delta $ of
Eqs. \eqref{du0=} and \eqref{Dv=vOm}

\begin{eqnarray}\label{var-u0=}
 i_\delta u_\mu^0 =u_\mu^I i_\delta \Omega^I\; , \qquad \delta  u_\mu^I =u_\mu^0 i_\delta  \Omega^I -  u_\mu^Ji_\delta \Omega^{JI} \; , \\ \nonumber \\ 
 \label{var-v=vivOm}
\delta v_\alpha{}^q
= \frac 1 2  i_\delta \Omega^I \gamma^I_{qp} v_\alpha{}^p - \frac 1 4 i_\delta \Omega^{IJ} v_\alpha{}^p\gamma^{IJ}_{pq}  \; . \qquad
\end{eqnarray}

When working with field theory in Harmonic superspace (which will happen when discussing the field theory resulting in D$0$--brane quantization) we have to use the covariant harmonic derivatives
\be\label{DIv}
\text{D}^I=\frac 1 2 v_\alpha^p \gamma^I_{pq}\frac {\partial} {\partial v_\alpha{}^q} \; , \qquad
\text{D}^{IJ}=\frac 1 2 v_\alpha^p \gamma^{IJ}_{pq}\frac {\partial} {\partial v_\alpha{}^q}\; . \qquad
\ee
These form the Lorentz group algebra
\be\label{alg=DIDJ}
{}[\text{D}^I,\text{D}^J]=\text{D}^{IJ}\; , \qquad
   {}[\text{D}^{IJ},\text{D}^K]=-2\delta^{K[I}\text{D}^{J]}\; , \qquad  {}[\text{D}^{IJ},\text{D}^{KL}]=-4\delta^{[K|[I}\text{D}^{J]|L]}\;  \qquad
\ee
and can be obtained by decomposing the differetial acting on $Spin(1,10)$ group manifold on Cartan forms,
\be\label{dv=OmD}
\text{d}= \text{d}v_\alpha{}^q \frac {\partial} {\partial v_\alpha{}^q} = \Omega^I \text{D}^I -
\frac 1 2 \Omega^{IJ}\text{D}^{IJ}\; .
\ee

When acting on vector harmonics (moving frame vectors) only, the following equivalent representation of the harmonic covariant derivatives can be used:
\be\label{DIu}
\text{D}^I=u_\mu^I \frac {\partial}{\partial u_\mu^0} + u_\mu^0 \frac {\partial}{\partial u_\mu^I} \; , \qquad
\text{D}^{IJ}=u_\mu^I \frac {\partial}{\partial u_\mu^J} -u_\mu^J \frac {\partial}{\partial u_\mu^I} \; . \qquad
\ee

\subsection{D$0$-brane in the analytical coordinate basis}
\label{D0=anatyt}

One of the advantages of the model with spinor moving frame variables is the possibility to change the coordinate basis replacing the spinor and vector coordinate functions by Lorentz invariant functions. Such a coordinate basis is called analytical \cite{Sokatchev:1985tc,Sokatchev:1987nk,Bandos:1990ji},  due to  historical reasons\footnote{This refers to that one of the roots of spinor moving frame method is Lorentz harmonic superspace approach to ${\cal N}=2$ and ${\cal N}=3$  supersymmetric theories in $\text{D}=4$ \cite{Galperin:1984av,Galperin:1984bu,Galperin:2001seg}.}. In our case, the coordinates of a suitable  analytical basis of the Lorentz harmonic superspace $\Sigma^{(10+45|16+16)}$
\be\label{LH=SSP=An}
{\cal Z}^{{\cal M}}_{An}= ({\rm x}^0, {\rm x}_{_{An}}^I, \Theta_q , \tilde{\Theta}_q, v_\alpha{}^q)
\ee
are related to the central basis coordinates  \eqref{LH=SSP} by

\bea\label{xAn0=}
{\rm x}^0= x^\mu u_\mu^0  \; , \qquad {\rm x}_{_{An}}^I= x^\mu u_\mu^I -\frac i 2\, (\theta^{q1}- \theta_q^{2})\gamma^I_ {qp}  (\theta^{p1}+ \theta_p^{2})  \; , \qquad \\ \nonumber \\ 
\label{ThAn=}
 \Theta^q = \theta^{q1}+ \theta_q^{2} \; , \qquad \tilde{\Theta}{}^q= \theta^{q1}- \theta_q^{2} \; , \qquad
 \theta^{q1}= \theta^{\alpha 1}v_\alpha^q\; , \qquad  \theta_q^{2}= \theta_\alpha^{2}v^\alpha_q\; . \qquad \\ \nonumber 
\eea
The expressions in \eqref{xAn0=} are chosen in such a way that  $\tilde{\Theta}{}^q$ disappears from the Lagrangian which reads

\be\label{cLD0=analyt}
{\cal L}_1^{{\rm D}0}=m \left(\text{d}{\rm x}^{{\rm 0}}- i\text{d}\Theta_q \, \Theta_q -\frac i 4 \Omega^{IJ}
\Theta\gamma^{IJ}\Theta - {\rm x}_{_{An}}^I \Omega^{I}\right)\; , \qquad \Theta\gamma^{IJ}\Theta\equiv \Theta_q\gamma^{IJ}_{qp}\Theta_p  \;
\ee
and includes only 16 of 32 fermionic coordinate functions,

\be
\Theta_q= \theta^{1q}+ \theta^2_q= \theta^{\alpha 1}v_\alpha{}^q +\theta_{\alpha}^{2}v_q{}^\alpha\;.
\ee
Notice that just this combination is invariant under the $\kappa$-symmetry transformations \eqref{eq:kappaD0=}. The same is true for $
{\rm x}_{_{An}}^I$ defined in \eqref{xAn0=}.

This implies that the local fermionic $\kappa$-symmetry is ``automatically'' fixed by passing to this basis.
Indeed, the only variable in the action \eqref{cLD0=analyt} which is not inert under $\kappa$-symmetry is
${\rm x}^0$ defined in  \eqref{xAn0=},

\be\label{dkx0=}\delta_\kappa {\rm x}^0 = i\kappa^q\Tilde{\Theta}_q \equiv i\kappa^q(\theta^{q1}-\theta_q^2)\; .
\ee
However this coordinate enters the Lagrangian only inside the total derivative term and hence
does not contribute to the action variation.

Furthermore, since entering the Lagrangian form under total derivative only, the coordinate function $ {\rm x}^0$ can actually be omitted from the action and from the configuration space of the model. After that, one can also speak about automatic gauge fixing of the reparametrization symmetry by passing to the analytical basis. We, however, prefer to keep the total derivative term $m \text{d} {\rm x}^0$ in the Lagrangian form since this simplifies spacetime interpretation of the field theory resulted from the D$0$-brane quantization.


To vary the action with the Lagrangian form \eqref{cLD0=analyt}, one has to take into account that
the Cartan forms, constructed from the moving frame variables as in Eqs. \eqref{OmI:=},
obey the Maurer-Cartan equations

\bea
{\cal D} \Omega^{I} &=& \text{d} \Omega^{I} + \Omega^{J}\wedge \Omega^{JI}=0 \; , \qquad  \\ \nonumber \\ \frak{F}{}^{IJ} &=& \text{d} \Omega^{IJ}+ \Omega^{IK}\wedge \Omega^{KJ}=-  \Omega^{I}\wedge \Omega^{J}\; \; .  \\ \nonumber
\eea
Then the variations of the Cartan forms can be obtained by formal application of Lie derivative formula,
$\delta = i_\delta \text{d} +\text{d}  i_\delta $ with identification of the independent variations with
$i_\delta   \Omega^{I} $ and $i_\delta   \Omega^{IJ} $. Notice that the latter are parameters of the  $\text{SO}(9)$ group
which is the (almost manifest) gauge symmetry of the Lagrangian form \eqref{cLD0=analyt}. The corresponding transformations of the Cartan forms and of the coordinate functions are

\bea
\delta_{\text{SO}(9)} \Omega^I=   i_\delta  \Omega^{IJ}\Omega^{J}\; , \qquad \delta_{\text{SO}(9)} \Omega^{IJ}= {\cal D} i_\delta\Omega^{IJ}=\text{d} i_\delta\Omega^{IJ}+i_\delta\Omega^{IK}\Omega^{KJ}-\Omega^{IK}i_\delta\Omega^{KJ}
   \;  , \qquad
\\
\delta_{\text{SO}(9)}  {\rm x}_{_{An}}^I =  i_\delta \Omega^{IJ}{\rm x}_{_{An}}^J\; , \qquad  \delta_{\text{SO}(9)} \Theta_q= \frac 1 4 i_\delta \Omega^{IJ} \gamma^{IJ}_{qp}\Theta_p  \; . \qquad
\eea

Under the essential variations, which correspond to coset $\text{SO}(1,9)/\text{SO}(9)$ and are parametrized  by $i_\delta   \Omega^{I} $, the Cartan forms transform as

\be
\delta \Omega^I= {\cal D} i_\delta \Omega^{I}=\text{d}i_\delta \Omega^{I} + i_\delta  \Omega^{J}\wedge \Omega^{JI} \; , \qquad \delta \Omega^{IJ}=  i_\delta \Omega^{I} \Omega^{J}-  i_\delta \Omega^{J} \Omega^{I} \; . \qquad
\ee
These relations allow us to find the equations of motion for admissible variations of the Cartan forms which are
\bea \label{eq:idOmI}
&& \text{D}{\rm x}_{_{An}}^I -\frac i 2 \Theta\gamma^{IJ}\Theta\, \Omega^J=0 \; , \qquad  \text{D}{\rm x}_{_{An}}^I :=\text{d}{\rm x}_{_{An}}^I -\Omega^{IJ}{\rm x}_{_{An}}^J\; , \\ \label{eq:idOmIJ}
&& {\rm x}_{_{An}}^{[I}\Omega^{J]}- \frac i 2 \Theta\gamma^{IJ}\text{D}\Theta=0\; , \qquad  \text{D}\Theta= \text{d}\Theta-\frac 1 4 \Omega^{IJ}\gamma^{IJ} \Theta\; . \qquad
\eea
Notice that to obtain \eqref{eq:idOmIJ} we have to use the conditions of the SO$(9)$ invariance  of $\gamma^I_{pq}$ which implies $ \Theta\gamma^{IJ}\text{D}\Theta= \Theta\gamma^{IJ}\text{d}\Theta+\Omega^{K[I} \Theta\gamma^{J]K}\Theta$.

The variation with respect to the fermionic coordinate function, $\delta \Theta_q$, results in vanishing of the covariant derivative of this function,
\be \label{eq:vTh}
\text{D}\Theta_q= \text{d}\Theta_q+\frac 1 4 \Omega^{IJ}\Theta_p\gamma^{IJ}_{pq}=0 \;  \qquad \ee
while the variation with respect to the coordinate function ${\rm x}_{_{An}}^I$ gives
\be \label{eq:vxI}\Omega^I=0 \; . \ee
Due to these two equations, Eq.~\eqref{eq:idOmIJ} is satisfied identically and this statement can be recognized as  the Noether identity for the $\text{SO}(9)$ gauge symmetry of D$0$-brane in the analytical coordinate basis.

Finally, as the coordinate function ${\rm x}^0$ enters the Lagrangian under the total derivative,  it produces only a trivial equation of motion. It is instructive, however, to write this trivial equation
of motion in a bit more detail:
\be
\text{d}m=0 \; .\ee
Indeed it is trivially satisfied as long as $m$ is a constant. However, this suggests the possibility of a generalization of the above Lagrangian for D$0$-brane in an analytical coordinate basis of Lorentz harmonic superspace  where constant $m$ is replaced by a function $m(\tau)$.

Clearly, the equations of such a model coincide with the original equations obtained above so that the only the difference will be that the value of constant $m$ will be fixed on-shell only, as a value of integration constant (as also happens with the cosmological constant in Unimodular gravity \cite{Unruh:1988in,Alvarez:2015pla} and 3-form (super)gravity \cite{Aurilia:1980xj,Ogievetsky:1980qp,Brown:1987dd,Ovrut:1997ur}).

\subsection{Hamiltonian formalism in the analytical basis}

In the models with spinor moving frame variables it is convenient to define the Legendre trasform from Lagrangian to canonical Hamiltonian by (see \cite{Bandos:2007wm}, \cite{Bandos:2024hnj} and refs. therein for details)

\bea\label{H0=}
\text{d}\tau H_0&=& \text{d}{\cal Z}^{{\cal M}}_{{\rm An}} {\cal P}_{{\cal M}} + \Omega^I \frak{d}^I - \frac 1 2 \Omega^{IJ} \frak{d}^{IJ}-{\cal L}_1^{{\rm D}0}\; .
\eea
This implies the standard definition of canonical momenta for usual the coordinate functions

\be\label{PM:=} {\cal P}_{{\cal M}}:= (p_{{\rm 0}}, p_{{\rm I}}, \Pi_{q}, \tilde{\Pi}_{q}) = \frac {\partial L} {\partial \dot{{\cal Z}}^{{\cal M}}_{{\rm An}}}\equiv \frac {\partial {\cal L}_1} {\partial \text{d}{\cal Z}^{{\cal M}}_{{\rm An}}}
\ee
as well as  the definition of covariant momenta for spinor moving frame variables are derivatives of the Lagrangian form with respect to Cartan forms,

\be\label{fdI:=}
\frak{d}^I = \frac  {\partial L}  {\partial \Omega^I_\tau}= \frac  {\partial {\cal L}_1}  {\partial \Omega^I}\; , \qquad \frak{d}^{IJ} =- \frac  {\partial L}  {\partial \Omega^{IJ}_\tau}= -\frac  {\partial {\cal L}_1}  {\partial \Omega^{IJ}}\; . \qquad
\ee

The Poisson--Dirac brackets on such a phase space  is defined by the standard 

\bea
{} [  {\cal P}_{{\cal M}}\, , {\cal Z}^{{\cal N}}_{{\rm An}}\}_{_{{\rm PB}}} = - (-)^{\epsilon ({\cal M})\, \epsilon ({\cal N})}  [  {\cal Z}^{{\cal N}}_{{\rm An}}\, ,  P_M\}_{_{{\rm PB}}} =-\delta_{{\cal M}}^{{\cal N}}\; , \qquad   \nonumber \\  \nonumber \\ {} [ {\cal Z}^{{\cal M}}_{{\rm An}}\, , {\cal Z}^{{\cal N}}_{{\rm An}}\}_{_{{\rm PB}}} =0\; , \qquad   [  {\cal P}_{{\cal M}}\, ,  {\cal P}_{{\cal N}}\}_{_{{\rm PB}}} =0\; , \qquad 
\eea
supplemented by
\be
{} [\frak{d}^I,  \ldots \}_{_{{\rm PB}}}  = -\text{D}^I \ldots \; ,  \qquad [\frak{d}^{IJ},  \ldots \}_{_{{\rm PB}}}  = -\text{D}^{IJ} \ldots \; ,
\ee
where multidots denote a function of moving frame vectors or spinor moving frame variables and $\text{D}^{I}$, $\text{D}^{IJ}$ are the covariant  derivatives defined in \eqref{DIu} and \eqref{DIv}.

Calculating the canonical momenta for coordinate functions (see Eq.~\eqref{PM:=}) and covariant momenta for the moving frame variables (see Eq.~\eqref{fdI:=}),  we arrive at the following set of primary constraints:

\bea
\label{p0-m=0}&& p_{{\rm 0}}-m:= p_{_{{\rm x}^0}}-m\approx 0 \; , \qquad  \\ \nonumber  \\ 
\label{pI=0}
&& p_{{\rm I}}:= p_{{\rm x}_{{\rm An}}^{{\rm I}}}\approx 0 \; , \qquad  \\ \nonumber  \\ \label{fdI-mxI=0}
&& \tilde{\frak{d}}^{I} ={\frak{d}}^{I} +m {\rm x}_{_{An}}^I  \approx 0 \; ,\qquad  \\ \nonumber  \\ \label{fdIJ-ThgIJTh=0}
&& \tilde{\frak{d}}^{IJ} ={\frak{d}}^{IJ}- \frac {im} 4 \Theta_q\gamma^{IJ}_{qp}  \Theta_p \approx 0 \; , \qquad  \\ \nonumber \\   \label{Pi+imTh=0} && \frak{d}_q:=\frak{d}_{_{\Theta_q}} = \Pi_q+ im\Theta_q \approx 0 \; . \qquad \\ \nonumber
\eea
It is easy to see that the canonical Hamiltonian \eqref{H0=} vanishes on the surface of these constraints.

The classification and analysis of the D$0$-brane constraints in the analytical basis is very simple (as suggested by previous experience with simpler systems, see in particular  \cite{Bandos:2023web}).
The bosonic constraints \eqref{fdI-mxI=0} and \eqref{pI=0} are clearly of the second class and, furthermore, are actually resolved

\be
{\rm x}_{_{An}}^I \approx -{\frak{d}}^{I}/m\; ,\qquad p_{{\rm I}}\approx 0
\ee
with respect to the pair of canonical variables ${\rm x}_{_{An}}^I , p_{{\rm I}}$ which therefore
can be removed from the phase space of the system \cite{Dirac:1963}. The bosonic constraint
\be
\tilde{\frak{d}}^{IJ} ={\frak{d}}^{IJ} +\frac 1 4 \Theta_q\gamma^{IJ}_{qp}  \Pi_p \approx 0 \; \qquad
\ee
obtained as a linear combination of \eqref{fdIJ-ThgIJTh=0} and \eqref{Pi+imTh=0} is of the first class. It generates $\text{SO}(9)$ gauge symmetry on the reduced phase space.

The last bosonic constraint \eqref{p0-m=0} is also of the first class. It generates local translations in ${\rm x}^0$ direction so that the field ${\rm x}^0(\tau)$ can be  gauged away. This reflects the fact that it enters the action under total derivative only and thus can be omitted from the phase space from the very beginning. As we have already mentioned, we  prefer to maintain this pure gauge variable as it allows us to keep track of the spacetime (central coordinate basis) origin of our model.

The fermionic constraints  \eqref{Pi+imTh=0}

\be\label{fdq:=} \frak{d}_{_{\Theta_q}}=: \frak{d}_q= \Pi_q+ im\Theta_q\quad (= \Pi_q+ i\Theta_q p_{_{{\rm x}^0}}) \ee
are imaginary and of the second class as they obey the algebra

\be
{}\{\frak{d}_q , \frak{d}_p\}_{_{\text{PB}}} = -2im \delta_{qp}\qquad (=-2i p_{_{{\rm x}^0}}\delta_{qp})\; .
 \ee
Here and below we will supplement the expressions involving $m$ by expressions with momentum
$p_{{\rm 0}}\equiv p_{{\rm x}^0}$ obtained with the use of constraint \eqref{p0-m=0}. This will be especially suggestive after  performing the quantization as then the substitution $m\mapsto -i\partial_{x^0}$ will often help to resolve possible doubts about the Hermiticity properties of the expressions.

\subsection{Quantization of the D$0$-brane: a way to field theory}
\label{QunatumD0}
In quantum theory the classical canonical variables are replaced by operators acting on state vectors. Roughly speaking (see discussion below), in the coordinate representation the state vector will be described by a superfield
\be\label{Xi=state}
\Xi ({\rm x}^{0}, v_\alpha{}^q, \Theta_q) \;  \qquad
\ee
and momenta are represented by differential operators

\bea
p_{{\rm 0}}&\mapsto & -i \partial_{{0}} \equiv -i \partial_{{\rm x}^0} \; , \qquad \\ \nonumber \\  
\Pi_q &\mapsto & -i \partial_q \equiv -i \partial_{_{\Theta_q}}  \; , \qquad \\ \nonumber \\  
\frak{d}^I &\mapsto & -i \text{D}^I \; , \qquad \\ \nonumber \\  
\frak{d}^{IJ} &\mapsto & -i \text{D}^{IJ} \; , \qquad 
\eea
where $\text{D}^I$ and $\text{D}^{IJ}$ are defined in \eqref{DIv}.

The state vector superfield can carry some indices,
\be\label{XicA=state}
\Xi_{{\cal A}} ({\rm x}^{0}, v_\alpha{}^q, \Theta_q) \; , \qquad
\ee
that is one has a set of superfields, probably of different statistics.  We will discuss this possibility below, but let us  not concentrate on it at the present stage.

According to Dirac's prescription \cite{Dirac:1963}, we have to impose the first class constraints on the state vector. Imposing the generator of $\text{SO}(9)$ gauge symmetry we arrive at the equation
\be
\left(\text{D}^{IJ} + \frac 1 4 \Theta_q\gamma^{IJ}_{qp}  \partial_{p}\right)\Xi=0 \qquad
\ee
which implies that the state vector superfield $\Xi$ is $\text{SO}(9)$ invariant. Imposing the last bosonic constraint  $\hat{p}_0- m$,
\be\label{id0-m}
(-i\partial_{{\rm x}^0}-m)\Xi =0~,
\ee
we fix the ${\rm x}^0$ dependence of the state vector to be exponential,
\be
\Xi ({\rm x}^0, v, \Theta)=
 e^{im{\rm x}^0 }\Phi (v, \Theta) \; . \qquad
\ee

The main problem is how to impose the quantum version of the fermionic constraints \eqref{fdq:=}.
Here we will follow
the method developed for  case of massless 10D and 11D
superparticle, which was elaborated
in \cite{Bandos:2017eof} and \cite{Bandos:2017zap}. It
begins by observing that the coordinate representation of the quantum version of the constraints \eqref{fdq:=} is given by
\be\label{fdq=-iDq} \hat{\frak{d}}_q= -i\text{D}_q \ee
where
\bea\label{Dq:=IIAm}
\text{D}_q=  \partial_{_{\Theta_q}} +i \Theta_q \partial_{{\rm x}^{\rm 0}}\; , \qquad
\eea
obeying
\be\label{DqDp=-2mI}
\{\text{D}_q,\text{D}_p\}=2i \delta_{qp}  \partial_{{\rm x}^{\rm 0}}\; .
\ee

%
\label{Quant=sec2}
As we have already mentioned, we cannot impose second class constraints \eqref{fdq=-iDq} on the state vector in the simplest manner,
i.e., by stating that state vector is in the kernel of this operator (as such a step would result in contradiction, with the expectation that the state vector is not equal to zero).

In the similar problem for massless 11D superparticle quantization
two approaches
are elaborated in  \cite{Bandos:2017eof,Bandos:2017zap}. One way is to allow the state vector to be multicomponent (see \eqref{XicA=state}), i.e. to be represented by several components with different $\text{SO}(9)$ indices and appropriate statistics, and try to realize the supersymmetry algebra of covariant derivatives on it by imposing some equation with nonvanishing r.h.s. \cite{Bandos:2017eof} (see \cite{Green:1999by,Bandos:2006nr} for related methods in the light-cone gauge and 11D generalization of twistor formalism respectively).

Another way, which we will take in our present study of D$0$-brane quantization, starts by introducing the complex structure and using it to split real 16-component  fermionic constraints in
a pair of 8 complex constraints and its c.c. To this end we can use the $\text{SO}(9)$ spinor frame formalism of \cite{Bandos:2017zap} the basic elements of which are rectangular blocks of the   $\text{Spin}(9)$ valued matrix
\begin{eqnarray}\label{winSpin=bww}
&& w_p^{(q)}= \left(\begin{matrix} \bar{w}_{pA} \;  w_p^A \end{matrix}\right) \in \text{Spin}(9)\; .
\end{eqnarray}
These obey the constraints

\begin{eqnarray}
\label{wbw=I}
&& \bar{w}_{qA}\bar{w}_{qB}=0 =w_q^A w_q^B\; , \qquad \bar{w}_{qA} w_q^B=\delta_A{}^B\; , \qquad \bar{w}_{qA} w_p^A +  w_q^A \bar{w}_{pA} = \delta_{qp}  \; ,  \qquad
\end{eqnarray}
which reflect the fact that $\text{Spin}(9)\subset \text{SO}(16)$, and provides a double covering of the $\text{SO}(16)$ frame

\begin{eqnarray}
\label{UinSO9} &&
  U_I^{(J)}= \left(U_I{}^{\check{J}}, \frac 1 2 \left( U_I+ \bar{U}_I\right), \frac 1 {2i} \left( U_I- \bar{U}_I \right)\right) \; \in \; \text{SO}(9)
 \;   \qquad
  \end{eqnarray}
in the sense of the following constraints

\begin{eqnarray} \label{wgIw=UcU}
  && \bar{w}_{qA}\gamma^{I}_{qp}\bar{w}_{pB}= U_I {\cal U}_{AB}\; , \qquad
  {w}_{q}^{A}\gamma^{I}_{qp}{w}_{p}^{B}= \bar{U}_I \bar{{\cal U}}^{AB}\; , \qquad
   \bar{w}_{qA}\gamma^{I}_{qp}{w}_{p}^{B}= iU_I ^{\check{J}}(\sigma^{\check{J}}\bar{{\cal U}})_{A}{}^{B}\;  , \qquad  \\ \nonumber \\ 
   \label{UIgI=}
&&  U_I \gamma^I_{qp} = 2\bar{w}_{qA} \bar{{\cal U}}{}^{AB} \bar{w}_{pB}\; , \qquad
\bar{U}{}_I \gamma^I_{qp} = 2{w}_{q}^{A} {{\cal U}}_{AB}w_p^{B} \; , \qquad
  U_I^{\check{J}} \gamma^I_{qp} =
i {w}_{q}^{A} (\sigma^{\check{J}}\bar{{\cal U}}){}_A{}^{B }\bar{w}_{pB} +i \bar{w}_{qA} (\tilde{\sigma}{}^{\check{J}}{\cal U})^A{}_B{w}_{p}^{B}\; .  \qquad
\end{eqnarray}
Here
 \be\label{tsI=-sI}
\tilde{\sigma}{}^{\check{J}AB}= -\sigma^{\check{J}}_{AB}=\; \sigma^{\check{J}}_{BA}=(\tilde{\sigma}{}^{\check{J}AB})^\dagger\;
\ee
are $\text{SO}(7)$ Clebsh-Gordan coefficients and  $ {\cal U}_{AB}$ is complex symmetric unitary matrix playing the role of a charge conjugation matrix for $\text{SO}(7)$. This and its c.c.  $\bar{{\cal U}}{}^{AB}=({\cal U}_{AB})^*$ obey

\be\label{UbU=1}
  {\cal U}_{AC}\bar{{\cal U}}{}^{CB}=\delta_A{}^B\; , \qquad {\cal U}_{AB}={\cal U}_{BA} \;, \qquad \bar{{\cal U}}{}^{AB}=\bar{{\cal U}}{}^{BA}=({\cal U}_{AB})^*\;  \qquad
\ee
and can be chosen to be unity matrices, ${\cal U}_{AB}=\delta_{AB}$,  $\bar{{\cal U}}{}^{AB}= \delta^{AB}$. We nevertheless prefer to keep for them separate notation (in particular, for  convenience, but also in relation with hidden $SU(8)$ symmetry in Appendices~\ref{sec:SU8inL} and \ref{SU8-SO7}).

Eq. \eqref{UinSO9} implies that the complex conjugate vectors $U_I$ and $\bar{U}_I$ are null and orthogonal to the remaining 7 vectors, 

\begin{eqnarray}\label{UU=0=} && U_IU_I=0=  \bar{U}_I \bar{U}_I \; , \qquad U_I \bar{U}_I=2 \; , \qquad U_IU_I{}^{\check{J}} =0=  \bar{U}_I U_I{}^{\check{J}} \; , \qquad  U_I{}^{\check{J}}U_I{}^{\check{K}}  =\delta^{\check{J}\check{K}}\; . \qquad \end{eqnarray}
See \cite{Bandos:2017eof,Bandos:2017zap} for further details.

At the present stage, Eqs.~\eqref{wbw=I} are most important for us. This implies that in the rest of this (sub)section we can work with SO(16) coordinates which obey the constraints
\eqref{wbw=I} only. This is reflection of hidden SU(8) symmetry ($SU(8)\subset SO(16)$) which we discuss in Appendix \ref{sec:SU8inL}. The restriction to $Spin(9)$ coordinates  by imposing constraints \eqref{wgIw=UcU}--\eqref{UIgI=} will be important in the next section where we will address the problem of spacetime interpretation of the on-shell superfield which we are going to obtain now as a representation of a quantum state vector of the D$0$-brane.

Eqs.~\eqref{wbw=I} allow us to replace an imaginary 16-component covariant derivative \eqref{Dq:=IIAm} obeying \eqref{DqDp=-2mI} by  complex octuplet of covariant derivatives and their (minus) complex conjugates

\bea\label{DA=bwD}
{\text{D}}_A &=&\bar{w}_{qA} \text{D}_q=\bar{\partial}{}^A + i \Theta^A\partial_{x^0}= \bar{\partial}{}^A \,  e^{- i \Theta^A\bar{\Theta}_A\partial_{x^0}} \; ,    \; , \qquad \nonumber \\   \nonumber \\ \bar{\text{D}}{}^A &=&  w_q^A \text{D}_q = {\partial}_A + i \bar{\Theta}_A\partial_{x^0}= {\partial}_A \,  e^{i \Theta^A\bar{\Theta}_A\partial_{x^0}} =-(\bar{\text{D}}{}^A)^*  \; , \qquad
\eea
which obey
\be
\{{\text{D}}_A, {\text{D}}_B\}=0\; , \qquad \{{\text{D}}_A, \bar{\text{D}}{}^B\}=-2m\delta_A{}^B \; , \qquad \{\bar{\text{D}}{}^A, \bar{\text{D}}{}^B\}=0\; .  \qquad
\ee
In \eqref{DA=bwD} we have introduced the complex 8-component coordinates and derivatives 

\bea\label{ThetaA:=}
&& \Theta^A= \Theta^qw_q^A\; , \qquad \bar{\Theta}_A= \Theta^q\bar{w}_{qA}\; ,  \qquad
\bar{\partial}{}^A =\frac {\partial}{\partial \bar{\Theta}_A} =- (\partial_A)^*\; . \eea

One of two complex conjugate quantum constraints  \eqref{DA=bwD} can be consistently imposed on the state vector thus realizing a Gupta--Bleuler like quantization prescription.
Then the state vector can be represented by the complex superfield $\Xi=\Xi (x^0, \Theta^A, \bar{\Theta}_A; v, w)$
which obeys
\be \label{bDAXi=0}
\bar{\text{D}}{}^A \Xi =0 \; .
\ee

A more rigorous method is to introduce the internal harmonics $w, \bar{w}$ already at the level of Lagrangian, to construct the Hamiltonian mechanics of the model in the configuration space enlarged in such a way, and then to perform quantization of such a Hamiltonian mechanics. We have followed this method as described in Appendix~\ref{sec:SU8inL} and have shown that the results of quantization are exactly the same as we
obtained using the simplified method
in the main text below.

The solution of Eq.~\eqref{bDAXi=0} can be written in the form of

\bea\label{Phi=phi+}
\Xi  (x^0, \Theta^A, \bar{\Theta}_A; v, w) &=& e^{-im(x^0+i\Theta^A\bar{\Theta}_A)}   \left(  \phi+ \Theta^A \psi_A + \frac 1 2 \Theta^B\Theta^A \phi_{AB} +
\frac 1 {3!} \Theta^C \Theta^B \Theta^A \psi_{ABC} + \right. \qquad  \nonumber \\  \nonumber \\   && + \frac 1 {4!} \Theta^D  \Theta^C \Theta^B \Theta^A \psi_{ABCD} +  \nonumber \\ \nonumber \\  
&& +\frac 1 {5!\, 3!} \epsilon_{A_1\ldots A_5B_1B_2B_3} \Theta^{A_5}  \ldots \Theta^{A_1} \tilde{\psi}{}^{B_1B_2B_3} +   \frac 1 {6!\, 2!} \epsilon_{A_1\ldots A_4B_1B_2} \Theta^{A_6}  \ldots \Theta^{A_1} \tilde{\phi}{}^{B_1B_2} + \qquad   \nonumber \\ \nonumber \\  
&& \left.  + \frac 1 {7!} \epsilon_{A_1\ldots A_7B} \Theta^{A_7}  \ldots \Theta^{A_1} \tilde{\psi}{}^{B} + \frac 1 {8!} \epsilon_{A_1\ldots A_8} \Theta^{A_8}  \ldots \Theta^{A_1} \tilde{\phi} \right) \; .  \qquad \\ \nonumber
\eea
This chiral superfield is similar to the so-called on-shell superfields for $\text{D}=4$ ${\cal N}=8$ supergravity theory \cite{ArkaniHamed:2008gz,Brandhuber:2008pf,Elvang:2015rqa}\footnote{These provide a basis to construct  superamplitude formalism, a superfield version of spinor helicity formalism related to the twistor approach.} and also to the chiral superfield that describes linearized type IIB supergravity \cite{Green:1982tk,Howe:1983sra}. The difference is in the type of moving frame variables the component fields depend on.

As in these cases, the chiral superfield \eqref{Phi=phi+} does not describe irreducible representation of supersymmetry (type IIA SUSY in our  case). Indeed, one can easily  reduce this representation by imposing the set of conditions expressing higher components of the superfields in terms of fields conjugate to its lower components,

\be\label{tpsi=psi*}
\tilde{\psi}{}^{B_1B_2B_3}\propto ({\psi}_{B_1B_2B_3})^*\; , \qquad \tilde{\phi}{}^{B_1B_2}\propto ({\phi}_{B_1B_2})^*\; , \qquad    \tilde{\psi}{}^{B} \propto ({\psi}_{B})^* \; ,  \qquad \tilde{\phi}\propto ({\phi})^*\; , \qquad
\ee
as well as the duality relations between the intermediate component and its c.c.,

\be
 \phi_{A_1\ldots A_4} =\frac 1 { 4!} \epsilon_{A_1\ldots A_4B_1B_2B_3B_4} \tilde{\phi}{}^{B_1B_2B_3B_4} \; , \qquad  (\phi_{B_1\ldots B_4})^* =\tilde{\phi}{}^{B_1B_2B_3B_4}\; .
\ee
These relations can be collected in one superfield duality equation between 4-th rank covariant spinor derivatives of the state vector superfield $\Xi$ and of its conjugate $\bar{\Xi}$

\bea\label{D4Xi=ebD4bXi}
 \text{D}_{A_1}\ldots \text{D}_{A_4}\Xi  =\frac 1 {4!} \epsilon_{A_1\ldots A_4B_1B_2B_3B_4} \bar{\text{D}}{}^{B_1}\ldots \bar{\text{D}}{}^{B_4}\bar{\Xi} \; , \qquad  (\Xi)^* =\bar{\Xi}\; . \\ \nonumber 
\eea

Of course, from the point of view of the quantization procedure, imposition of the additional condition \eqref{D4Xi=ebD4bXi} requires some explanation. It can be motivated by the requirement that the state vector should be described by an irreducible representation of supersymmetry, which in its turn can be followed back to the paradigm  that elementary particle is described by irreducible representation of the Poincar\'e group \cite{Novozhilov:1975yt}.

Imposing the condition \eqref{D4Xi=ebD4bXi}, we reduce the field content of the state vector superfield to complex bosonic scalar and second rank $\text{SO}(7)$ spin-tensor

\be
\phi (v,w,\bar{w})\; , \qquad \phi_{AB} (v,w,\bar{w})=- \phi_{BA} (v,w,\bar{w})\; , \qquad \phi_{A_1\ldots A_4} =\frac 1 {4!\, 4!} \epsilon_{A_1\ldots A_4B_1B_2B_3B_4}   (\phi_{B_1\ldots B_4})^*\;
\ee
and fermionic spinor and 3-rank antisymmetric spin--tensor

\be
\psi_A (v,w,\bar{w})\; , \qquad \psi_{ABC} (v,w,\bar{w})= \psi_{[ABC]} (v,w,\bar{w})\;  . \qquad
\ee

Experience with the supertwistor approach and on-shell superfields of $\text{D}=4$  ${\cal N}=8$ supergravity and $\text{D}=4$ ${\cal N}=4$  SYM theories \cite{ArkaniHamed:2008gz,Brandhuber:2008pf,Elvang:2015rqa} (extended to 11D SUGRA and 10D SYM in \cite{Bandos:2017zap,Bandos:2017eof}, see also  \cite{Galperin:1992pz}) suggests that the components of the on-shell superfield parametrize the general solutions of the equation of the corresponding supersymmetric field theory.
Our next problem is to identify which supermultiplet of type IIA supersymmetry appears as the quantum state spectrum of D$0$-brane as described by chiral superfield \eqref{Phi=phi+} obeying also \eqref{D4Xi=ebD4bXi}.

\setcounter{equation}0

\section{Quantum state spectrum of the D$0$-brane }
\label{QSpectrum}
The quantum state spectrum of the D$0$-brane should be a massive field theory with type IIA supersymmetry, maximal supersymmetry with 32 charges. It is clear that such a supermultiplet is associated with higher spin fields (appearing after its dimensional reduction to $\text{D}=4$), in particular massive spin 2 field, and one might expect a problem in constructing consistent interaction for such fields.  However, in the case of
the D$0$-brane we can expect that  any possible inconsistency when considering the interactions can be resolved in the complete description of type IIA string theory, the integral part of which is the D$0$-brane.
Notice also a consonance with recent interest to coupling of the massive spin 2 field to $\text{D}=4$ supergravity \cite{Bossard:2025jjw}.

Our strategy here is heuristic: to guess which massive type IIA supermultiplet should form the quantum state spectrum of D$0$-brane, to solve the linearized field equations  for such a supermultiplet in terms of our spinor frame variables and to show that the parameters of such a solution is in one to one correspondence with the field content of the chiral superfield \eqref{Phi=phi+} subject to the constraint \eqref{D4Xi=ebD4bXi}.

The natural suspicion is that the field theory which is obtained by quantization of D$0$-brane is a massive counterpart of the linearized type IIA supergravity. Furthermore, as type IIA supergravity can be obtained by dimensional reduction of the maximal 11D supergravity,  the natural candidate for its  massive counterpart is a massive mode of the Kaluza-Klein reduction of 11D supergravity.

\subsection{Equations for massive type IIA fermionic fields}

\subsubsection{11D Rarita-Schwinger equation and convenient dimensional reduction ansatz}
It is natural to begin with dimensional reduction of gravitino, the only fermionic field of 11D supergravity. This is described by a vector-spinor
\be
\underline{\Psi}_{\underline{\mu}}^{\underline{\alpha}}(\underline{x}^{\underline{\nu}})= \left(\underline{\Psi}_{{\mu}}^{\underline{\alpha}}(\underline{x}^{\underline{\nu}}) , \underline{\Lambda}^{\underline{\alpha}}(\underline{x}^{\underline{\nu}}) \right)\; \qquad
\ee
which obeys the 11D massless Rarita-Schwinger (RS) equation
\be\label{RS=11D}
\Gamma^{\underline{\mu}\underline{\nu}\underline{\rho}}_{\underline{\alpha}\underline{\beta}}\partial_{\underline{\nu}}
\underline{\Psi}{}_{\underline{\rho}}{}^{\underline{\beta}}=0\; .
\ee
Eq.~\eqref{RS=11D} is invariant under gauge supersymmetry in its linearized version
\be\label{susy=11D}
\delta \underline{\Psi}{}_{\underline{\nu}}{}^{\underline{\beta}}=
\partial_{\underline{\nu}}\underline{\epsilon}{}^{\underline{\beta}}\; .
\ee

Here and below the underlined symbols from the middle and from the beginning of the Greek alphabet denote the 11-vectors and 11D Majorana spinor indices, which can be also considered as Majorana spinor indices in 10D,
\bea
& \nonumber \underline{\mu},\underline{\nu},\underline{\rho}= 0,1,\ldots 9,10\; , \qquad \underline{\alpha}, \underline{\beta}, \underline{\gamma}=1,\ldots,32\; . \eea
Furthermore, we also underline  coordinates and the fields that depend on all 11 coordinates, so that
\bea\label{11->10}
& \underline{x}^{\underline{\nu}} =(x^\nu, y^*) \; , \qquad \underline{\partial}_{\underline{\nu}} =({\partial}_\nu , \partial_*)\; , \qquad  \underline{\Psi}_{{\mu}}^{\underline{\alpha}}=\left(\begin{matrix}\underline{\Psi}_{{\mu}}^{{\alpha}1}(x,y^*)\cr \underline{\Psi}_{\mu\alpha{}}^{2}(x,y^*)\end{matrix}\right) \; , \qquad  \underline{\Lambda}^{\underline{\alpha}}=\left(\begin{matrix}\underline{\Lambda}^{{\alpha}2}(x,y^*)\cr \underline{\Lambda}_{\alpha}{}^{1}(x,y^*)\end{matrix}\right)\; .
\eea
The seemingly strange distribution of the 1 and 2 superindices in the last equation is related, in particular, to the specific $y^*$ dependence of the fields in our Kaluza-Klein ansatz. Namely, to be consistent we  set

\bea\label{ansatz=1}
\underline{\Psi}_{{\mu}}^{{\alpha}1}(x,y^*) = {\Psi}_{{\mu}}^{{\alpha}1}(x)\, \cos (my^*)\; , \qquad \underline{\Lambda}_{\alpha}{}^{1}(x,y^*) = {\Lambda}_{\alpha}{}^{1}(x) \cos (my^*) \; , \qquad
\\  \label{ansatz=2}
 \underline{\Psi}_{\mu\alpha{}}^{2}(x,y^*)={\Psi}_{\mu\alpha{}}^{2}(x)\,  \sin (my^*)   \; , \qquad  \underline{\Lambda}^{{\alpha}2}(x,y^*)= {\Lambda}^{{\alpha}2}(x) \,  \sin (my^*)   \; . \qquad
\eea

The simplest way to obtain  the equations for massive type IIA fields is through the gauge fixing of the local supersymmetry by imposing the gamma-tracelessness condition
\bea\label{GPsi=0}
& \Gamma^{\underline{\mu}}_{\underline{\alpha}\underline{\beta}}
\underline{\Psi}{}_{\underline{\mu}}{}^{\underline{\beta}}=
\Gamma^{{\mu}}_{\underline{\alpha}\underline{\beta}}
\underline{\Psi}{}_{{\mu}}{}^{\underline{\beta}}+\Gamma^{*}_{\underline{\alpha}\underline{\beta}}
\underline{\Lambda}{}^{\underline{\beta}}
=0\;
\eea
and reducing the  RS equations \eqref{RS=11D} to the Dirac equation and divergence-free or transversality condition

\bea
\label{Dirac=11D} \Gamma^{\underline{\nu}}_{\underline{\alpha}\underline{\beta}}\partial_{\underline{\nu}}
\underline{\Psi}{}_{\underline{\mu}}{}^{\underline{\beta}}=\Gamma^{{\nu}}_{\underline{\alpha}\underline{\beta}}\partial_{\underline{\nu}}
\underline{\Psi}{}_{{\mu}}{}^{\underline{\beta}}+ \Gamma^{*}_{\underline{\alpha}\underline{\beta}}\partial_{*}
\underline{\Lambda}{}^{\underline{\beta}}=0\; ,
\\
\label{div=0=11D} \partial^{\underline{\mu}}
\underline{\Psi}{}_{\underline{\mu}}{}^{\underline{\alpha}}=\partial^{{\mu}}
\underline{\Psi}{}_{{\mu}}{}^{\underline{\alpha}}-\partial_{*}
\underline{\Lambda}{}^{\underline{\alpha}}=0\; .
\eea

This set of equations is invariant under the residual  linearized supersymmetry \eqref{susy=11D} with the fermionic parameter obeying Dirac-Weyl equation
\be\label{susy=11D-}
\delta \underline{\Psi}{}_{\underline{\nu}}{}^{\underline{\beta}}=
\partial_{\underline{\nu}}\underline{\epsilon}{}^{\underline{\beta}} \; , \qquad \Gamma^{\underline{\nu}}_{\underline{\alpha}\underline{\beta}}\partial_{\underline{\nu}}\underline{\epsilon}{}^{\underline{\beta}}=0\; .
\ee

\subsubsection{Solving 11D Rarita-Schwinger equation with spinor moving frame method }
The Fourier image of the  above 11D equations  can be solved,  using the 11D spinor moving frame variables described below, in terms of the gamma-traceless $\text{SO}(9)$ spin-tensor $\Psi^I_{{q}}$ \cite{Galperin:1992pz}, by

\be\label{Psi=11D=uIvPsiI}
\underline{\Psi}{}_{\underline{\mu}}{}^{\underline{\alpha}}(v^-,v^+)= \underline{u}^I_{\underline{\mu}}v{}_{{q}}^{-\underline{\alpha}} \Psi^I_{{q}}\; , \qquad \gamma^{I}_{pq}\Psi^I_{q}=0\; , \qquad I=1,\ldots,9\; , \qquad p,q =1,\ldots,16\; , \qquad
\ee
where $\gamma^I_{qp}=\gamma^I_{p{q}}$ are $\text{SO}(9)$ Dirac matrices obeying \eqref{gIgJ=dIJ}.

In Eq.~\eqref{Psi=11D=uIvPsiI} $v{}_{{q}}^{-\underline{\alpha}}$ is 11D  the spinor moving frame variable  described below and
$\underline{u}_{\underline{\mu}}^{I}$ is a nanoplet of orthogonal and normalized 11-vectors which, together with two light-like vectors $\underline{u}_{\underline{\mu}}^{\#}$ and $ \underline{u}_{\underline{\mu}}^=$
form the $\text{SO}(1,9)$ valued {\it moving frame matrix}
\begin{eqnarray}\label{harmU=in11D}
 & \underline{u}_{\underline{\mu}}^{(b)}(\tau ) =
 \left( {1\over 2}\left(  \underline{u}_{\underline{\mu}}^{\#}+ \underline{u}_{\underline{\mu}}^=
 \right), \;  \underline{u}_{\underline{\mu}}^{I} \, ,
 {1\over 2}\left( \underline{u}_{\underline{\mu}}^{\#}
 -  \underline{u}_{\underline{\mu}}^=\right)\right)\; \in \; \text{SO}^\uparrow (1,10)\,  .
\end{eqnarray}
The name {\it moving frame matrix} reflects the  identification of light-like 11-momentum $p_\mu$  with one of the  light-like vectors involved in it. We chose this to be $\underline{u}_{\underline{\mu}}^=$,
\bea\label{p=r++u--=11D}
& \underline{p}{}_{\underline{\mu}}= \rho^\# \underline{u}_{\underline{\mu}}^=\; ,
\eea
and make the identification up to a multiplier $\rho^\#$. This is a St\"uckelberg field for $\text{SO}(1,1)$ symmetry necessary to streamline  the use of group-theoretical structure beyond the moving frame and spinor moving frame variables (see e.g.  \cite{Bandos:2007wm} and refs. therein).

The spinor moving frame matrix
\begin{eqnarray}\label{harmV==11D}
V_{\underline{\alpha}}^{(\underline{\beta})}= \left(\begin{matrix} v_{\underline{\alpha} {q}}^{\; +} , & v_{\underline{\alpha} q}^{\; -}
  \end{matrix}\right) \in \text{Spin}(1,10)\;  \qquad
\end{eqnarray}
provides a ``square root''  of the moving frame matrix
\eqref{harmU=in11D} in the sense that the following constraints hold:

\begin{eqnarray}\label{u==v-v-=11D}
 &
  v^-_{{q}} \tilde{\Gamma}_{{\underline{\mu}}}v^-_{{p}}=\underline{u}_{\underline{\mu}}^= \delta_{{q}{p}}   \; , & \qquad  \underline{u}_{\underline{\mu}}^= \Gamma^{\underline{\mu}}_{\underline{\alpha}\underline{\beta}}= 2v_{\underline{\alpha} q}{}^- v_{\underline{\beta} q}{}^-  \; ,  \qquad \\
\label{v+v+=u++}
&  \;  v_{{q}}^+ \tilde{\Gamma}_{{\underline{\mu}}} v_{{p}}^+ =\underline{u}_{{\underline{\mu}}}^{\# } \delta_{{q}{p}}\; , & \qquad \underline{u}_{ {\underline{\mu}}}^{\# } {\Gamma}^{ {\underline{\mu}}}_{ {{\underline{\alpha}}} {{\underline{\beta}}}} =2 v_{{{\underline{\alpha}}}{q}}{}^{+}v_{{{\underline{\beta}}}{q}}{}^{+} \; , \qquad \\
 \label{uIs=v+v-=11D}
& v_{{q}}^- \tilde {\Gamma}_{{\underline{\mu}}} v_{{p}}^+=\underline{u}_{ {\underline{\mu}}}^{I} \gamma^I_{q{p}}\; , &\qquad
 \underline{u}_{{\underline{\mu}}}^{I} {\Gamma}^{{\underline{\mu}}}_{{\underline{\alpha}}{\underline{\beta}}} =2 v_{( {{\underline{\alpha}}}|{q} }{}^- \gamma^I_{qp}v_{|{{\underline{\beta}}}){p}}{}^{+} \; . \quad  \end{eqnarray}
The solution \eqref{Psi=11D=uIvPsiI} also involves $v{}_{{q}}^{-\underline{\alpha}}$ which is the rectanguar block of the matrix inverse to the spinor frame matrix
\eqref{harmV==11D}, $
V_{(\underline{\beta})}^{\; \;\underline{\alpha}}= \left(\begin{matrix} v_q^{-\underline{\alpha} }, & v_q^{+\underline{\alpha}}
  \end{matrix}\right) \in \text{Spin}(1,10)$,
which implies
\begin{eqnarray}\label{VV-1=1}
v_q^{-\underline{\alpha} } v_{\underline{\alpha} {p}}^{\; +}=\delta_{qp}\; , \qquad
v_q^{-\underline{\alpha} } v_{\underline{\alpha} p}^{\; -} =0  \; , \qquad
v_q^{+\underline{\alpha}}v_{\underline{\alpha} {p}}^{\; +}=0  \; , \qquad  v_q^{+\underline{\alpha}}
 v_{\underline{\alpha} p}^{\; -}= \delta_{qp}\; . \qquad
\end{eqnarray}

Using  these equations and the following consequences of
Eq.~\eqref{harmU=in11D} 

\begin{eqnarray}\label{uu=011D}
&& \underline{u}_{\underline{\mu}}^=\underline{u}^{\underline{\mu}=}=0 \; , \qquad \underline{u}_{\underline{\mu}}^\#\underline{u}^{\underline{\mu}\#}=0 \; , \qquad \underline{u}_{\underline{\mu}}^=\underline{u}^{\underline{\mu}\#}=2 \; , \qquad   \underline{u}_{\underline{\mu}}^I\underline{u}^{\underline{\mu}\#}=0 \; , \quad 
\underline{u}_{\underline{\mu}}^I\underline{u}^{\underline{\mu}J}=-\delta^{IJ} \; , \qquad
\end{eqnarray}
we find the solution \eqref{Psi=11D=uIvPsiI} of the 11D Rarita-Shwinger equation\footnote{Actually, the solution contains the additional term $+u^=_{\underline{\mu}}v{}_{\underline{q}}^{-\underline{\alpha}} \Psi^{[+3]}_{\underline{q}}$, but this can be gauged away using the residual local supersymmetry \eqref{susy=11D-}.
}.

\subsubsection{Dimensional reduction and solution of equations for  massive type IIA gravitini}
To have the equations for 10D massive fermions, we have to use the Kaluza-Klein reduction ansatz \eqref{ansatz=1}, \eqref{ansatz=2},
the $\text{SO}(1,9)$ invariant decomposition  \eqref{11->10} and  $\text{SO}(1,9)$ invariant representation for 11D Dirac matrices

\bea\label{G11=s10}
&& \Gamma^\mu_{\underline{\alpha}\underline{\beta}} = \left(\begin{matrix} \sigma^{\mu}_{\alpha\beta} & 0\cr
 0 & \tilde{\sigma}{}^{\mu\alpha\beta}
\end{matrix}\right)\; , \qquad  \Gamma^*_{\underline{\alpha}\underline{\beta}} = \left(\begin{matrix} 0 & - \delta_{\alpha}{}^{\beta} \cr
- \delta_{\beta}{}^{\alpha}  & 0
\end{matrix}\right)\; , \qquad
\\ \label{tG11=s10}
&& \tilde{\Gamma}^{\mu \underline{\alpha}\underline{\beta}} = \left(\begin{matrix}  \tilde{\sigma}{}^{\mu\alpha\beta}& 0\cr
 0 & \sigma^{\mu}_{\alpha\beta}
\end{matrix}\right)\; , \qquad \; \tilde{\Gamma}^{*\underline{\alpha}\underline{\beta}} = \left(\begin{matrix} 0 & \delta_{\beta}{}^{\alpha} \cr
\delta_{\alpha}{}^{\beta} & 0
\end{matrix}\right)\; , \qquad
\\ \label{C11=10}
&& C_{\underline{\alpha}\underline{\beta}} = i\,\left(\begin{matrix} 0 &\delta_{\alpha}{}^{\beta} \cr
 -\delta_{\beta}{}^{\alpha}  & 0
\end{matrix}\right)\; , \qquad C^{\underline{\alpha}\underline{\beta}} = i\,\left(\begin{matrix} 0 & \delta_{\beta}{}^{\alpha} \cr
-\delta_{\alpha}{}^{\beta} & 0
\end{matrix}\right)\; . \qquad
\eea
Using these equations,  the straightforward calculations
allow one to find that Eq. \eqref{GPsi=0}, \eqref{Dirac=11D} and \eqref{div=0=11D} imply

\bea\label{sPsi=10D}
\sigma^{\mu}_{\alpha\beta}{\Psi}{}_{\mu}{}^{\beta 1}=\Lambda_{\alpha}^1\; &, \qquad  &
\tilde{\sigma}^{\mu \alpha\beta} {\Psi}{}_{\mu \beta}{}^2= {\Lambda}{}^{\alpha 2}\; , \qquad
\\
\label{Dirac=10DP}
{\partial}\!\!\!/{} _{\alpha\beta} {\Psi}{}_{\mu}{}^{\beta 1}=m{\Psi}{}_{\mu \alpha}{}^2\; & , \qquad &
 \tilde{\partial}\!\!\!/{}^{\alpha\beta}{\Psi}{}_{\mu \beta}{}^2=-m {\Psi}{}_{\mu}{}^{\alpha 1}\; , \qquad
\\
\label{Dirac=10DL}
 \tilde{\partial}\!\!\!/{}^{\alpha\beta} {\Lambda}{}_{\beta}{}^1=m {\Lambda}{}^{\alpha 2}\; &, \qquad &
{\partial}\!\!\!/{} _{\alpha\beta} {\Lambda}{}^{\beta 2}=-m{\Lambda}{}_{\alpha}{}^1\; , \qquad
\\
\label{div=10DPL} \partial^{\mu}
{\Psi}{}_{\mu}{}^{\alpha 1}= m {\Lambda}{}^{\alpha 2}\; &, \qquad &
\partial^{\mu}
{\Psi}{}_{\mu \alpha}{}^{ 2}=- m {\Lambda}_{\alpha}{}^1\;, \eea
where
\bea
{\partial}\!\!\!/{} _{\alpha\beta}=  \sigma^{\nu}_{\alpha\beta}\partial_\nu
\; , \qquad \tilde{\partial}\!\!\!/{}^{\alpha\beta} =
 \tilde{\sigma}^{\nu \alpha\beta}\partial_\nu
\; .
 \eea

Some simple observations to make are, first, that the Dirac equations \eqref{Dirac=10DP} and \eqref{Dirac=10DL} imply that both gravitini and dilatini obey the standard Klein-Gordon equation,

\be
(\Box +m^2) {\Psi}{}_{\mu}{}^{\alpha 1}=0 \; , \qquad (\Box +m^2){\Psi}{}_{\mu \alpha}{}^{ 2} =0 \; , \qquad (\Box +m^2) {\Lambda}{}_{\beta }^1=0  \quad {\text{and}} \quad (\Box +m^2) {\Lambda}{}^{\beta 2}=0 \qquad
\ee
with $\Box +m^2 :=\partial_\mu\partial^\mu +m^2$, second, that using \eqref{sPsi=10D} we can write Eqs.\eqref{div=10DPL} in terms of gravitini only,

\bea
\label{div=10DP} \partial^{\mu}
{\Psi}{}_{\mu}{}^{\alpha 1}= m \tilde{\sigma}^{\mu \alpha\beta} {\Psi}{}_{\mu \beta}{}^2\; &, \qquad &
\partial^{\mu}
{\Psi}{}_{\mu \alpha}{}^{ 2}=- m \sigma^{\mu}_{\alpha\beta}{\Psi}{}_{\mu}{}^{\beta 1}\;. \qquad
 \eea
The third observation is that  Eqs.~\eqref{sPsi=10D}-\eqref{div=10DPL} are invariant under the reduced supersymmetry \eqref{susy=11D-}
parametrized by
\bea\label{epsilon=IIAm}
\underline{\epsilon}^{{\alpha}1}(x,y^*) =
{\epsilon}\!\!\!^{{^0}}{}^{{\alpha}1}+{\epsilon}^{{\alpha}1}(x)\cos (my^*)\; , \qquad
\underline{\epsilon}_{\alpha{}}^{2}(x,y^*)={\epsilon}\!\!\!^{{^0}}{}_{\alpha{}}^{2}+{\epsilon}_{\alpha{}}^{2}(x) \sin (my^*)   \; . \qquad
\eea
Here ${\epsilon}\!\!\!^{{^0}}{}^{{\alpha}1}$ and  ${\epsilon}\!\!\!^{{^0}}{}_{\alpha{}}^{2}$ are constant parameters while the fermionic functions ${\epsilon}^{{\alpha}1}(x)$ and  ${\epsilon}_{\alpha{}}^{2}(x)$
obey massive Dirac equations

\bea\label{Dirac=10Dep}
{\partial}\!\!\!/{} _{\alpha\beta} {\epsilon}{}^{\beta 1}=m{\epsilon}{}_{\alpha}{}^2\; , \qquad \tilde{\partial}\!\!\!/{}^{\alpha\beta} {\epsilon}{}_{\beta}{}^2=-m {\epsilon}{}^{\alpha 1}\; .  \qquad
\eea

The ``zero modes'' of 11D supersymmetry parameter,  ${\epsilon}\!\!\!^{{^0}}{}^{{\alpha}1}$ and  ${\epsilon}\!\!\!^{{^0}}{}_{\alpha{}}^{2}$, do not contribute to the transformations of gravitini which obey free
equations. However, it is important to keep their presence in mind when supersymmetry of (massive) supergravity action or equations are considered: they  give rise to the supersymmetry \eqref{susy=hmn} of the linearized massive supergravity action \eqref{L=massiveIIA}.

In the case of gravitini obeying the free equations,  the  supersymmetry \eqref{epsilon=IIAm}  acts on the 10D fields as 

\bea\label{susy=IIAm1}
\delta{\Psi}_{{\mu}}^{{\alpha}1}(x)\, =\partial_\mu {\epsilon}^{{\alpha}1}(x) \; , \qquad
\delta  {\Lambda}_{\alpha}{}^{1}(x) = m{\epsilon}_{\alpha}{}^2(x)  \; , \qquad
 \\ \label{susy=IIAm2}
\delta {\Psi}_{\mu\alpha{}}^{2}(x)= \partial_\mu {\epsilon}_{\alpha}{}^2(x)   \; , \qquad  \delta  {\Lambda}^{{\alpha}2}(x) = -m{\epsilon}^{{\alpha}1}(x)  \; . \qquad
\eea
Thus massive dilatini transform as Volkov-Akulov Goldstonions.
Furthermore, as these are Goldstone fields for local supersymmetry, they can be gauged away by using the residual supersymmetry parameters obeying the same Dirac equations as the Goldstonions do. Thus we can fix the local supersymmetry by setting
\be \Lambda_{\alpha}{}^1 =0=\Lambda^{\alpha 2}\; . \ee
Then the independent part of the gauge fixed equations are

\bea
\label{Dirac=m-10DP}
{\partial}\!\!\!/{} _{\alpha\beta} {\Psi}{}_{\mu}{}^{\beta 1}=m{\Psi}{}_{\mu \alpha}{}^2\; & , \qquad &
 \tilde{\partial}\!\!\!/{}^{\alpha\beta}{\Psi}{}_{\mu \beta}{}^2=-m {\Psi}{}_{\mu}{}^{\alpha 1}\; , \qquad
\\
\label{sPsi=0-10D}
\sigma^{\mu}_{\alpha\beta}{\Psi}{}_{\mu}{}^{\beta 1}=0\; &, \qquad  &
\tilde{\sigma}^{\mu \alpha\beta} {\Psi}{}_{\mu \beta}{}^2= 0\; , \qquad
\\
\label{divP=0=10D} \partial^{\mu}
{\Psi}{}_{\mu}{}^{\alpha 1}= 0\; &, \qquad &
\partial^{\mu}
{\Psi}{}_{\mu \alpha}{}^{ 2}=0\;.  \eea

We however, find it instructive  to keep  the nonvanishing $\Lambda_{\alpha}{}^1$ and $\Lambda^{\alpha 2}$  and 10D gauge supersymmetry for some further stages
\footnote{
Notice, that in certain circumstances  allowing non-vanishing traces and gamma-traces of tensorial and spin-tensorial fields  significantly simplifies the study of a system under consideration. For example,  working with reducible representations of the Poincar\'e group allows to formulate linearized $\mathcal{N}=1$ supergravities
 along with linear and tensor multiplets
in $\text{D}=4,6$ and $10$ in a simple and uniform way \cite{Sorokin:2018djm}
 (see also \cite{Bertrand:2022pyi}).  }.

\subsection{Solutions of the equations for massive type IIA fermionic fields I: SO(9) covariant form}
A convenient way to pass to the momentum representation for the field equations is to consider real plane waves of the type

\bea
{\Psi}{}_{\mu}{}^{\alpha 1}(x)= {\Psi}{}_{\mu}{}^{\alpha 1}(p)\cos(p_\mu x^\mu) \; &, \qquad &
\Lambda_{\alpha}^1(x)= \Lambda_{\alpha}^1(p)\cos (p_\mu x^\mu) \; , \qquad   \\
{\Psi}{}_{\mu \alpha}{}^2(x) = -{\Psi}{}_{\mu \alpha}{}^2(p) \sin(p_\mu x^\mu) \; & , \qquad & {\Lambda}{}^{\alpha 2} (x)= -{\Lambda}{}^{\alpha 2}(p) \sin (p_\mu x^\mu)\; . \qquad
\eea
Then equations for the fields in the momentum representation are \eqref{sPsi=10D} and

\bea
\label{Dirac=10DP-p}
{p}\!\!\!/{} _{\alpha\beta} {\Psi}{}_{\mu}{}^{\beta 1} =m{\Psi}{}_{\mu \alpha}{}^2\; & , \qquad &
\tilde{p}\!\!\!/{}^{\alpha\beta} {\Psi}{}_{\mu \beta}{}^2=m {\Psi}{}_{\mu}{}^{\alpha 1}\; , \qquad
 \\
\label{Dirac=10DL-p}
\tilde{p}\!\!\!/{}^{\alpha\beta}  {\Lambda}{}_{\beta}{}^1=m {\Lambda}{}^{\alpha 2}\; &, \qquad &
{p}\!\!\!/{} _{\alpha\beta}  {\Lambda}{}^{\beta 2}=m{\Lambda}{}_{\alpha}{}^1\; , \qquad
 \\
\label{div=10DPL-p} p^{\mu}
{\Psi}{}_{\mu}{}^{\alpha 1}= m {\Lambda}{}^{\alpha 2}\; &, \qquad &
p^{\mu}
{\Psi}{}_{\mu \alpha}{}^{ 2}=m {\Lambda}_{\alpha}{}^1\;.\eea
where
\be
{p}\!\!\!/{} _{\alpha\beta}= p_\nu \sigma^{\nu}_{\alpha\beta}
\; , \qquad \tilde{p}\!\!\!/{}^{\alpha\beta} =
p_\nu \tilde{\sigma}^{\nu \alpha\beta}
\; . \qquad
\ee

The consistency of Eqs.~\eqref{Dirac=10DP-p} and \eqref{Dirac=10DL-p} requires the momentum to be timelike
\be\label{p2=m2}
p_\mu p^\mu =m^2 \; .
\ee
The above ansatz remains invariant under supersymmetry \eqref{susy=IIAm1}, \eqref{susy=IIAm2} with parameters defined by
\be
{\epsilon}{}^{\alpha 1} (x)= {\epsilon}{}^{\alpha 1}(p) \sin (p_\mu x^\mu)\; , \qquad
\epsilon_{\alpha}^2(x)= \epsilon_{\alpha}^2(p)\cos (p_\mu x^\mu) \; , \qquad
\ee
where ${\epsilon}{}^{\alpha 1}(p)$ and $
\epsilon_{\alpha}^2(p)$ obey

\be
\\
\label{Dirac=10Dep-p}
{p}\!\!\!/{} _{\alpha\beta}  {\epsilon}{}^{\beta 1}=m{\epsilon}{}_{\alpha}{}^2\; , \qquad
\tilde{p}\!\!\!/{}^{\alpha\beta}  {\epsilon}{}_{\beta}{}^2=m {\epsilon}{}^{\alpha 1}\;  . \qquad \ee
This supersymmetry acts on the fields in momentum representation as follows:
\bea\label{susy==IIAm1}
\delta{\Psi}_{{\mu}}^{{\alpha}1}(p)\, = p_\mu {\epsilon}^{{\alpha}1}(p) \; , \qquad
\delta  {\Lambda}_{\alpha}{}^{1}(p) = m{\epsilon}_{\alpha}{}^2(p)  \; , \qquad
 \\ \label{susy==IIAm2}
\delta {\Psi}_{\mu\alpha{}}^{2}(p)= p_\mu {\epsilon}_{\alpha}{}^2(p)   \; , \qquad  \delta  {\Lambda}^{{\alpha}2}(p) = m{\epsilon}^{{\alpha}1}(p)  \; . \qquad
\eea

Let us solve Eq.~\eqref{p2=m2} in terms of spinor moving frame  \eqref{v=inSpin},
\begin{eqnarray}\label{ps==mvv}
&& p_\mu \sigma^\mu_{\alpha\beta}= m v_\alpha{}^q v_\beta{}^q \; ,   \qquad  m v_\alpha{}^q  \tilde{\sigma}_\mu^{\alpha\beta}v_\beta{}^p=p_\mu \delta_{pq} \;  .  \end{eqnarray}
Then the Dirac equations \eqref{Dirac=10DL-p} for two dilatino fields are solved by
\be
\Lambda_\alpha^1= v_\alpha{}^q \Lambda_q \; , \qquad \Lambda^{\alpha 2}= v_q{}^\alpha \Lambda_q \; , \qquad
\ee
in terms of the single fermionic 16 component spinor of $\text{SO}(9)$, $\Lambda_q$. Similarly, the Dirac equations \eqref{Dirac=10DL-p} for gravitini is solved by
\be
\Psi_{\mu}^{\alpha 1}= v_q{}^\alpha  \Psi_{\mu q} \; , \qquad \Psi_{\mu \alpha}{}^2= v_\alpha{}^q \Psi_{\mu q} \; , \qquad
\ee
and \eqref{div=10DPL-p} implies that
\be
\Psi_{\mu q} =u_\mu^0\Lambda_q - u_\mu^I\Psi^I_q \; ,
\ee
where $u_\mu^0$ and $u_\mu^I$ are moving frame vectors \eqref{harmU=m} which, let us recall, are related to spinor moving frame \eqref{v=inSpin} by \eqref{u0s=vv}.

Finally we have to substitute these expressions into Eqs.~\eqref{sPsi=10D} which gives us $\text{SO}(9)$ gamma-tracelessness condition for $\Psi^I_{q}$,
\be\label{giPiq=0}
\gamma^I_{pq}\Psi^I_{q}=0 \; . \qquad
\ee

Thus we conclude that the fermionic equations of the massive counterpart of type IIA supergravity theory (massive Kaluza-Klein mode of the 11D supergravity dimensionally reduced to 10D) is expressed in terms of the
$\text{SO}(9)$ fermionic spinor $\Lambda_q$ and of the $\text{SO}(9)$ vector-spinor $\Psi_{q}^I$ obeying the gamma-tracelessness conditions \eqref{giPiq=0}.

However, we still have the residual local supersymmetry with parameters obeying \eqref{Dirac=10Dep-p}. This can be  solved by $
\epsilon_\alpha^2= v_\alpha{}^q \epsilon^q$,  $\epsilon^{\alpha 2}= v_q{}^\alpha \epsilon^q $
and act on the fields  by
\be
\delta \Lambda_q=  m \epsilon^q  \; , \qquad \delta  \Psi^I_{q}=0\; .
\ee

Thus we can fix the gauge $\Lambda =0$ and reduce the spectrum of the solution to the only one gamma-traceless $\text{SO}(9)$ vector-spinor
\be\label{PsiSo9-IIAm}
\Lambda_{q}=0\; , \qquad  \Psi^I_{q}\; , \qquad \gamma^I_{pq}\Psi^I_{q}=0\; , \qquad I=1,\ldots,9\; , \qquad q,p=1,\ldots,16\; .
\ee

\subsection{Solutions of the equations for massive type IIA fermionic fields II: SO(7) covariant form}

The next stage is to decompose the irreducible representations of SO(9) carried by this gamma-traceless vector-spinor field into the irreducible representations of $\text{SO}(7)$. To this end, we use the $\text{SO}(9)$ spinor frame formalism of \cite{Bandos:2017zap} reviewed in sec.~\ref{Quant=sec2}. Particularly, Eqs.~\eqref{UinSO9}-\eqref{wgIw=UcU} allow us to decompose the $\text{SO}(9)$ vector-spinor  \eqref{PsiSo9-IIAm} into

\bea \Psi^I_q&=&\frac 1 2 U_I \bar{w}_{qA} \Psi^{(\bar{U})A} + \frac 1 2 U_I  w_{q}{}^A  \bar{\Psi}{}^{(\bar{U})}_A +\frac 1 2 \bar{U}_I \bar{w}_{qA} \Psi^{(U)A} + \frac 1 2 \bar{U}_I  w_{q}{}^A  \bar{\Psi}{}^{(U)}_A +  +~U_I^{\check{J}} \bar{w}_{qA} \Psi^{\check{J}A} +  U_I^{\check{J}}  w_{q}{}^A  \bar{\Psi}{}^{\check{J}}_A
\eea
and to split the 9d gamma-tracelessness condition \eqref{giPiq=0} onto

\be\label{sPsi=Psi}
\tilde{\sigma}{}^{\check{J}AB}  \bar{\Psi}{}{}^{\check{J}}_{B}=-i  {\Psi}{}{}^{(U)A}\; ,\qquad
\sigma{}^{\check{J}}_{AB}  {\Psi}{}{}^{\check{J}B}=-i \bar{\Psi}{}{}^{(\bar{U})}_{A}\; . \qquad
\ee
Eqs.~\eqref{sPsi=Psi} can be used to determine ${\Psi}{}{}^{(U)A}$
and $\bar{\Psi}{}{}^{(\bar{U})}_{A}$ in terms of  $\Psi^{\check{J}A}$ and  $\bar{\Psi}{}^{\check{J}}_A$, so that these latter fields
can be considered to be unrestricted.
Thus, the fermionic field content of the on-shell massive IIA supergravity is described by the following unrestricted fields

\bea\label{L+Psi=SO7}
\Psi^{(\bar{U})A}\; , \; \bar{\Psi}{}^{({U})}_A=(\Psi^{(\bar{U})A})^* \qquad {\rm and} \qquad  \Psi^{\check{J}A}\; , \;  \bar{\Psi}{}^{\check{J}}_A  =(\Psi^{\check{J}A})^*
\; , \qquad   \check{J}=1,\ldots,7\; , \qquad A,B=1,\ldots,8\; . \qquad
\eea

Now let us observe that  $\bar{\Psi}{}^{\check{J}}_A$ are $8\times7=56$ complex fields, and the same number of degrees of freedom  are enclosed in the  complex field $\psi_{ABC}=\psi_{[ABC]}$ which enters the on-shell chiral  superfield \eqref{Phi=phi+} obtained when quantizing the
D$0$-brane. The relation can be explicitly written as
\be \bar{\Psi}{}^{\check{J}}_A = \tilde{{\sigma}}{}^{\check{J} BC}\psi_{ABC} \; . \ee
The remaining complex fermionic $ \bar{\Psi}{}^{(\bar{U})}_A$
field can be identified with the $\psi_A$ component of the on-shell superfield,
\be \bar{\Psi}{}^{(\bar{U})}_A =\psi_{A}\; . \ee

Thus we conclude that the spectrum of fermionic fields contained in the decomposition \eqref{Phi=phi+}, \eqref{tpsi=psi*} of the  on-shell superfield obeying \eqref{bDAXi=0}  and \eqref{D4Xi=ebD4bXi}, which was  obtained by quantizing the D$0$-brane, can be identified with ``parameters'' of the solution of the equations for massive type IIA vector-spinor fermionic field which can be identified with the nontrivial mode of the dimensional  reduction of 11D gravitino, 

\bea \Psi^I_q&=&\frac 1 2 U_I  w_{q}{}^A  \psi_A +\frac 1 2 \bar{U}_I \bar{w}_{qA} \bar{\psi}^{A} +  U_I^{\check{J}} \bar{w}_{qA}{\sigma}^{\check{J}}_{BC}\bar{\psi}^{ABC}  +  U_I^{\check{J}}  w_{q}{}^A \tilde{{\sigma}}{}^{\check{J} BC}\psi_{ABC}
\\   \nonumber \\   \nonumber && (128=8+8+56+56)\; .
\eea

\subsection{Equations for massive  bosonic fields of massive type IIA supermultiplet and their solutions}
\label{Eqs=m=fixed}
Let us consider the bosonic sector of the eleven dimensional supergravity, which consists of the graviton
$h_{\underline{\mu} \underline{\nu}}$ and
the three-form gauge field $A_{\underline{\mu} \underline{\nu} \underline{\rho}
}=A_{[\underline{\mu} \underline{\nu} \underline{\rho}
]}$. We are going to perform the
Kaluza-Klein type dimensional reduction to
ten dimension by taking the first non-zero mode in decomposition on the 11-th coordinate $y^*$. In this way we shall obtain the bosonic superpartners  of the massive fermionic fields, obtained in the previous subsection.

The linearized Riemann tensor reads
\be\label{R=pph}
 R_{{\underline \mu } {\underline \nu}{\underline \rho}{ \underline \sigma}} = -4 \partial_{[\underline \mu |}  \partial_{[\underline{\rho} } h_{{\underline \sigma}]{|\underline \nu}]}
\ee
and the linearized Einstein equation obeyed by 11D graviton,
\be\label{Ei=EqR}
R^{\underline \nu}{}_{{\underline \mu_1 } {\underline \nu} {\underline \mu_2}} =  0\; ,
\ee
can be written in more detail as
\be \label{l-gr}
\Box \, h_{{\underline \mu_1} {\underline \mu_2}}-
\partial_{\underline{\mu_1} }
\partial^{\underline{\nu}} \, h_{\underline{\nu} \underline{\mu_2}}
-
\partial_{\underline{\mu_2} }
\partial^{\underline{\nu}}
\,
h_{\underline{\nu} \underline{\mu_1}}
+ \partial_{\underline{\mu_1} }
\partial_{\underline{\mu_2}} \,
h^{\underline{\nu}}{}_{\underline{\nu}}
=0 \; .
\ee
In this latter form  it is (just a bit) less trivial to prove
the invariance  under the diffeomorphism transformations,
\be \label{gauge-gr}
\delta \, h_{\underline{\mu_1} \underline{\mu_2}}=
\partial_{\underline{\mu_1} } \xi_{\underline{\mu_2} }+
\partial_{\underline{\mu_2} } \xi_{\underline{\mu_1} }\; ,
\ee
which is manifest in the original form  \eqref{Ei=EqR} with \eqref{R=pph}. However, the reduction to the Klein-Gordon equation  after imposing the so-called harmonic gauge $\partial^{\underline{\mu}} h_{\underline{\mu}\underline{\nu}}=\frac 1 2 \partial_{\underline{\nu}} h^{\underline{\mu}}{}_{\underline{\mu}} $ is more obvious in it.

As this gauge in its turn is invariant under the residual symmetry with parameter $\xi^\mu$ in \eqref{gauge-gr} obeying the Klein-Gordon equation, this transformation can be further used to set to zero the trace of the 11D graviton.
This additional  gauge fixing makes the linearized equation  for 11D supergravity graviton \eqref{Ei=EqR} equivalent to the set ($\underline{\Box}:=\partial^{\underline{\nu}}\partial_{\underline{\nu}}$)
\be
\underline{\Box} h_{\underline{\mu} \underline{\nu}}=0\; , \qquad  \partial^{\underline{\mu}} h_{\underline{\mu} \underline{\nu}}=0 \; , \qquad  h^{\underline{\mu}}{}_{\underline{\mu} }=0 \; . \qquad
\ee

This set of equations is still preserved by diffeomorphisms \eqref{gauge-gr} with the parameter obeying both Klein-Gordon and divergence-free conditions
\be
\underline{\Box} \underline{\xi}^{\underline{\mu}}=0 \; , \qquad \partial_{\underline{\mu}}\underline{\xi}^{\underline{\mu}}=0\;
\ee
which must be taken into account when writing the complete gauge fixed solution.

Passing to the momentum representation,  we find that
the Fourier image $h_{\underline{\mu} \underline{\nu}}(\underline{p})$ of the  11D graviton field $h_{\underline{\mu} \underline{\nu}}(\underline{x})$ is ``localized'' on the light-like momenta, i.e. on the mass shell, $\underline{p}{}_{\underline{\mu}}\underline{p}^{\underline{\mu}}=0$,
and obeys
\be\label{ph=0=trh}
\qquad  \underline{p}^{\underline{\mu}} h_{\underline{\mu} \underline{\nu}}(\underline{p})=0 \; , \qquad  h^{\underline{\mu}}{}_{\underline{\mu} }(\underline{p})=0 \; ,  \qquad \underline{p}{}_{\underline{\mu}}\underline{p}^{\underline{\mu}}=0\, . \qquad
\ee
This set of equations is invariant under Fourier image of diffeomorphisms \eqref{gauge-gr}
\be\label{diff-p}
\delta  h_{\underline{\mu} \underline{\nu}}(\underline{p}) =
\underline{p}_{\underline{\mu} } \underline{\xi}_{\underline{\nu} }+
\underline{p}_{\underline{\nu} } \underline{\xi}_{\underline{\mu} }\; ,
\ee
with parameter obeying
\be\label{pxi=0} \underline{p}^{\underline{\mu} } \underline{\xi}{}_{\underline{\mu} }=0
\; , \qquad \underline{p}_{\underline{\mu}}\underline{p}^{\underline{\mu}}=0\, . \qquad
\ee

Attaching the 11D moving frame \eqref{harmU=in11D}  to the light-like 11-momentum by \eqref{p=r++u--=11D}, we can write the solution of
Eq. \eqref{ph=0=trh} in terms of symmetric traceless
$\text{SO}(9)$ tensor $h_{IJ}$:

\be\label{h=uIuJh}
 h_{\underline{\mu} \underline{\nu}}(\underline{p})=u^I_{\underline{\mu}}u^J_{\underline{\nu} }
 h^{IJ} (\rho^\# ,\underline{u}^=) , \qquad
 h^{IJ}=h^{JI}\; , \qquad h^{II}=0\; . \qquad
\ee
To arrive at this form of the solution we have used the residual diffeomorphism symmetry \eqref{diff-p} after  solving the conditions \eqref{pxi=0} for its parameter by
$
 \underline{\xi}{}_{\underline{\mu}}(\underline{p})=u^=_{\underline{\mu}}\underline{\xi}^\# + u^I_{\underline{\nu} }
 \underline{\xi}^{I} $.

Dimensional reduction implies splitting 11D graviton field in the blocks

\be
 h_{\underline{\mu} \underline{\nu}}({\underline{x}})=\left(\begin{matrix}\underline{h}_{\mu\nu}
 & \underline{A}_\nu \cr
 \underline{A}_\mu & \underline{h}\end{matrix}\right)
 \; ,
\qquad
\ee
where the underline indicates a dependence on all 11 coordinates ${\underline{x}}^{\underline{\mu}}$,
and imposing the ansatz for the dependence of fields on the $y^*$ coordinate. We chose the following  dimensional reduction ansatz:

\be\label{ansatz=h}
\underline{h}_{\mu\nu}(x, y^*) = h_{\mu\nu}(x) \cos(my^*)\; , \qquad  \underline{A}_{\mu}(x, y^*) = A_{\mu}(x) \sin(my^*)\; , \qquad  \underline{h}(x, y^*) = h(x) \cos(my^*)\;  \qquad
\ee
This keeps 10D fields real and  implies that they satisfy the Klein-Gordon equation with mass $m$,

\be
(\Box +m^2) h_{\mu\nu}(x)=0\; , \qquad (\Box +m^2) A_{\mu}(x)=0\; , \qquad (\Box +m^2) h(x)=0\; .  \qquad
\ee
Furthermore, the 10D fields obey
\bea\label{hmumu=h}
\partial^\mu h_{\mu\nu}(x)=mA_\nu\;  , \qquad \partial^\mu A_{\mu}(x)=-mh\;  , \qquad h_\mu{}^\mu=h\; .  \qquad
\eea

Instead of trying to solve these equations as they are, we would like first to fix the gauge using the residual diffeomorphism symmetry. Its parameter compatible with the ansatz \eqref{ansatz=h}
split onto ten-vector and 10D scalar components dependent differently on $y^*$:
\bea\label{ansatz=xi}
& \underline{\xi}{}_{\mu}(x, y^*)= \xi_\mu(x) \cos(my^*)\; , \qquad  \underline{\xi}{}_{*}(x, y^*)= \xi_*(x) \sin(my^*)\; , \qquad
\eea
with functions in the r.h.s.-s obeying
\be
(\Box+m^2)\xi_\mu =0\; , \qquad
 (\Box+m^2)\xi_* =0\; , \qquad \partial^\mu \xi_\mu = m\xi_* \; .
\ee
Clearly, the last equation makes $\xi_*$ dependent  so  the only free parameter  that remains is
$\xi_\mu $. As both fields and parameters satisfy the massive Klein-Gordon equation, this can be used to fix the gauge $A_\mu=0$. In this gauge the second equation in \eqref{hmumu=h} gives $h=0$ so that we finally have only one symmetric  traceless divergent free tensor field  obeying the massive Klein-Gordon equation,

\bea\label{hmumu=0} 
(\Box +m^2) h_{\mu\nu}(x)=0\; , \qquad
\partial^\mu h_{\mu\nu}(x)=0\;  , \qquad  h_\mu{}^\mu=0\; .  \qquad
\eea

\subsection{The solutions of the equations for massive bosonic fields of type IIA supermultiplet}
Making the Fourier transform and solving Eqs.~\eqref{hmumu=0} in terms of a moving frame \eqref{harmU=m} attached to the momentum by
\be\label{p=mu0}
p_\mu = m u_\mu^{{\rm 0}}\; , \ee
we find
\be\label{h=uIuJh=m}
 h_{\mu\nu}= u_\mu^I u_\nu^J h^{IJ}\; , \qquad
 h^{IJ}=h^{JI}\; , \qquad h^{II}=0\; .
\ee
So, like in the 11D case (see \eqref{h=uIuJh}), the solution is expressed in terms of a symmetric and traceless $\text{SO}(9)$ tensor of second rank, but now this tensor field depends on 10D vector frame variables, \eqref{harmU=m}, $h^{IJ}= h^{IJ}(u_\mu^{{\rm 0}})$.

Now the problem is to find the expression for the symmetric traceless $\text{SO}(9)$ tensor  $h^{IJ}$ in terms of spin-tensor bosonic fields $\phi_{AB}$, $\bar{\phi}^{AB}=(\phi_{AB})^*$,
$\phi_{ABCD}= \frac 1 {4!} \epsilon_{ABCDEFGH}\bar{\phi}{}^{EFGH}$ and scalars
$\phi$, $\bar{\phi}=(\phi)^*$, which, as we have shown,  describe the bosonic sector of the quantum state spectrum of the D$0$-brane. This can be done by
using the $\text{SO}(9)/[\text{SO}(7)\otimes \text{SO}(2)]$ vector harmonics ($\text{SO}(9)$ moving frame) and the properties of the $\text{SO}(7)$ sigma matrices presented in Appendix~\ref{SU8-SO7}.
The study described in Appendix~\ref{hIJ-AIJK=SO7} results in the following expression 

\bea\label{hIJ=SO7}
h_{IJ} &=& U_I \, U_J \phi + \bar{U}_I \, \bar{U}_J \bar{\phi} +
U^{\check{I}}_{(I} U_{J)} \tilde{\sigma}{}^{\check{I}AB}\phi_{AB}+U^{\check{I}}_{(I}  \bar{U}_{J)} {\sigma}{}^{\check{I}}_{AB}\bar{\phi}^{AB}
 +~U^{\check{I}}_{I} U^{\check{J}}_{J}
\tilde{t}{}_{\check{I}\check{J}}^{ABCD}\phi_{ABCD}- \frac 1 2 U_I \, \bar{U}_J\tilde{t}{}_{\check{J}\check{J}}^{ABCD}\phi_{ABCD} \;   \nonumber \\  \nonumber
\\ && (44=1+1 +7+7 +28)\; , \eea  \nonumber
where
\bea
 \label{t24:=SO7=} t^{\check{I}\check{J}}_{ABCD}&=&
(\sigma^{\check{I}\check{K}}{\cal U})_{[AB}(\sigma^{\check{J}\check{K}}{\cal U})_{CD]} + \sigma^{\check{I}}{}_{[AB}\sigma^{\check{J}}{}_{CD]} \; . \qquad \eea

To check that the r.h.s. of \eqref{hIJ=SO7} is real, one has to take into account that $t^{\check{I}\check{J}}_{ABCD}$ is related to its c.c.

\bea  \label{tt24:=SO7=} \tilde{t}{}_{\check{I}\check{J}}^{ABCD}&=& (\tilde{\sigma}^{\check{I}\check{K}}\bar{{\cal U}})^{[AB}(\tilde{\sigma}^{\check{J}\check{K}]}\bar{{\cal U}})^{CD]}+\tilde{\sigma}^{\check{I}[AB}\tilde{\sigma}^{CD]\check{J}}=(t^{\check{I}\check{J}}_{ABCD})^*\;
\eea
by duality relation
\bea
 \label{t24=ebt24=}
\tilde{t}{}_{\check{I}\check{J}}^{ABCD}= \; \frac 1 {4!}  \epsilon^{ABCDEFGH}
t^{\check{I}\check{J}}_{EFGH} \; .
\eea

Similarly, one can show that the gauge fixed version of the Maxwell equation for 3-form gauge field

\be \label{3form-e=}
\partial^{\underline{\mu}}
F_{\underline{\mu} \underline{\nu} \underline{\rho} \underline{\sigma}}=0, \qquad
F_{\underline{\mu} \underline{\nu} \underline{\rho} \underline{\sigma}} =
\partial_{\underline{\mu}}\,
A_{ \underline{\nu} \underline{\rho} \underline{\sigma}}-
\partial_{\underline{\nu}}\,
A_{  \underline{\rho} \underline{\sigma} \underline{\mu}} +
\partial_{\underline{\rho}}\,
A_{  \underline{\sigma} \underline{\mu} \underline{\nu}}-
\partial_{\underline{\sigma}}\,
A_{  \underline{\mu} \underline{\nu} \underline{\rho}}
\ee
splits into the Klein-Gordon equation and  divergence-free conditions
\be \label{BoxA3=11D}
\underline{\Box} A_{ \underline{\mu} \underline{\nu} \underline{\rho}}=0\; , \qquad \partial^{ \underline{\rho}} A_{ \underline{\mu} \underline{\nu} \underline{\rho}}=0\; , \qquad
\ee
which in momentum representation are solved by
\be
{ A}_{ \underline{\mu} \underline{\nu} \underline{\rho}} = u^{I}_{ \underline{\mu}  }u^{J}_{\underline{\nu}  }u^{K}_{\underline{\rho}}  \underline{ A}_{IJK} \;
\ee
in terms of the antisymmetric $\text{SO}(9)$ tensor  $ \underline{ A}_{IJK}=  \underline{ A}_{[IJK]}
(\rho^\#, u^{=}_{ \underline{\mu}  })$ and moving frame vectors
$u^{I}_{ \underline{\mu}}$  orthogonal to the 11D momentum (see   \eqref{harmU=in11D} and \eqref{uu=011D}).

Furthermore, one can show that the dimensionally reduced gauge fixed equations for 3-form gauge field are given by the massive Klein-Gordon equation and 10D divergenceless condition

\be \label{BoxA3=10D}
({\Box}+m^2) A_{ {\mu}{\nu} {\rho}}=0\; , \qquad \partial^{ {\rho}} A_{ {\mu} {\nu} {\rho}}=0\; . \qquad
\ee
These equations are solved in terms of a similar antisymmetric $\text{SO}(9)$ tensor, that depends on a timelike 10-momentum $A_{IJK}= A_{[IJK]}
(u^{{\rm 0}}_{\mu})$.

The expression for this on-shell field in terms of spin-tensor  and scalar bosonic fields describing the (bosonic sector of the) quantum state spectrum of D$0$-brane is derived in Appendix~\ref{hIJ-AIJK=SO7}. Here we present only the result, which is
\bea \label{AIJK=SO7}
A_{IJK} &=&
U^{\check{I}}_{[I} U_{J} \bar{U}_{K ]} \; i\tilde{t}_{\check{I}}^{ABCD}
\, \phi_{ABCD}
+  U^{\check{I}}_{[I} U^{\check{J}}_{J} {U}^{\check{K}}_{K ]} i\tilde{t}_{\check{I}\check{J}\check{K}}^{ ABCD}
\, \phi_{ABCD} +
 \nonumber  \\ \nonumber  \\
&& +~U^{\check{I}}_{[I} {U}^{\check{J}}_{J} {U}_{K]} \,
(\tilde{\sigma}{}^{\check{I}\check{J}}\bar{{\cal U}})^{AB} \phi_{AB} +
 U^{\check{I}}_{[I} {U}^{\check{J}}_{J} \bar{U}_{K]}\, ({\sigma}{}^{\check{I}\check{J}}{\cal U})_{AB} \,
\bar{\phi}^{AB}
 \; , \qquad  \nonumber \\  \nonumber  \\ && {}\qquad (84=7+35 +21+21) \; ,  \eea
where
\bea\label{t34:=SO7=}
t^{\check{I}\check{J}\check{K}}_{ABCD}&=&
(\sigma^{[\check{I}\check{J}}{\cal U})_{[AB}\sigma^{\check{K}]}{}_{CD]}
\; , \qquad \\  \nonumber  \\ \label{tt34:=SO7=} \tilde{t}{}_{\check{I}\check{J}\check{K}}^{ABCD}&=& (\tilde{\sigma}^{[\check{I}\check{J}|}\bar{{\cal U}})^{[AB}\tilde{\sigma}^{CD]\, |\check{K}]}=(t^{\check{I}\check{J}\check{K}}_{ABCD})^*= -\frac 1 {4!}  \epsilon^{ABCDEFGH}
t^{\check{I}\check{J}\check{K}}_{EFGH} \; , \qquad \\ \nonumber  
\\ \label{t14:=SO7=} t^{\check{I}}_{ABCD}&=&
-(\sigma^{[\check{I}\check{K}}{\cal U})_{[AB}\sigma^{\check{K}]}{}_{CD]}\; , \qquad \\  \nonumber  \\ \label{tt14:=SO7=} \tilde{t}{}_{\check{I}}^{ABCD}&=&\; (\tilde{\sigma}^{\check{I}\check{K}}\bar{{\cal U}})^{[AB}\tilde{\sigma}^{}{}^{CD]\, \check{K}}=(t^{\check{I}}_{ABCD})^*=\frac 1 {4!}  \epsilon^{ABCDEFGH}
t^{\check{I}}_{EFGH}\; . \qquad \\ \nonumber  
\eea
See Appendix~\ref{SU8-SO7} for properties of these tensors -- spin-tensors of $\text{SO}(7)$ group and  other technical  details.

In the last lines of Eqs.~\eqref{hIJ=SO7} and \eqref{AIJK=SO7}
we have presented the counting of the balance of degrees of freedom in the l.h.s. and r.h.s. There the number of components taken from spin-tensor fields are calculated from $\text{SO}(7)$ vector indices of the $\text{SO}(7)$-invariant  tensors -- spin-tensors they are contracted with. The redistribution of the degrees of freedom of the spin-tensor fields on these $\text{SO}(7)$ tensors is described in Appendix~\ref{SU8-SO7}. It is not difficult to check that all the components of these spin-tensor fields are present in the r.h.s.-s of  Eqs.~\eqref{hIJ=SO7} and \eqref{AIJK=SO7}.

Therefore,  bosonic degrees of freedom contained in
the 10D massive fields $h_{\mu \nu}$ and $A_{\mu \nu \rho}$
have been  ``distributed'' among the scalar and spin-tensor fields $\phi$, $\phi_{AB}$ and $\phi_{ABCD}$.

\subsection{Linearized equations of the D$0$-brane field theory in 10D spacetime, its action and supersymmetry}
To conclude, let us write the complete covariant form of the equations for the fields of the massive counterpart of the type IIA supergravity in 10D  which, as we have shown,  are the (linearized) equations of the D$0$-brane field theory. These can be obtained by  dimensional reduction of the covariant form of the linearized 11D supergravity equations \eqref{Ei=EqR} (equivalent to \eqref{l-gr}), \eqref{3form-e=} and  \eqref{RS=11D}.

With our ansatz \eqref{ansatz=h} the linearized Einstein equation splits into three components, with two $*$-indices, with one $*$ index and with two 10D indices which give, respectively, equations with the Klein-Gordon operator acting on the
scalar, vector and tensor field,

\bea\label{Rmu*mu*=0}
 R^{\rho}{}_{*\; \rho *}= 0 \qquad &\Longrightarrow&    \qquad
 \Box h-m\partial^{\nu}A_{\nu}- m^2  h_\nu{}^{\nu}=0\; , \qquad  \\ \nonumber \\ \label{Rnumunu*=0}
R^{\rho}{}_{\mu\; \rho *}= 0 \qquad &\Longrightarrow&    \qquad \Box A_{\mu}+ m\partial^{\rho}h_{\rho\mu}-\partial_{\mu}\partial^{\rho}A_{\rho} -m \partial_{\mu}h_\rho{}^{\rho}=0\; , \qquad 
\\ \nonumber \\ \label{l-gr-1}  R^{\rho}{}_{\mu\; \rho \nu} - R_{*\mu\; * \nu}= 0
  \qquad &\Longrightarrow&  \qquad
(\Box \, + m^2) h_{{ \mu} { \nu}}-
2\partial_{({\mu} }
\partial^{{\rho}} \, h_{{\nu}) {\rho}}
+ \partial_{{\mu} }
\partial_{{\nu}} \,
(h^{{\rho}}{}_{{\rho}}-h) +2m \partial_{(\mu } A_{\nu)}
=0
\; . \\ \nonumber
\eea

Obviously, the vector and scalar fields, $A_\mu$ and $h$, are pure gauge or St\"uckelberg fields with respect to diffeomorphism transformations \eqref{gauge-gr} with parameters obeying \eqref{ansatz=xi}. This is to say, this can be used to fix the gauge
\be \label{A1=0=h} A_{\mu}=0\; ,  \qquad h=0\; . \ee
In this gauge \eqref{Rmu*mu*=0} and  \eqref{Rnumunu*=0}
 imply vanishing of trace and of divergence of   $h_{\mu \nu}$, respectively, while \eqref{l-gr-1} reduces to the massive Klein-Gordon equation for $h_{\mu \nu}$. In such a way we arrive at a gauge fixed description of  sec.~\ref{Eqs=m=fixed}, Eqs.~\eqref{hmumu=0}.

It is known, Eqs.~\eqref{hmumu=0}, which can be obtained from \eqref{Rmu*mu*=0}-\eqref{l-gr-1} by gauge fixing St\"uckelberg fields to zero,
can also be obtained from a unique Fierz-Pauli equation

\bea\label{Fierz-Pauli}
(\Box +m^2) h_{\mu\nu}-2 \partial^{\rho} \partial_{(\mu}h_{\nu )\rho}+ \partial_{\mu}\partial_{\nu}h_{\rho}{}^\rho  +\eta_{\mu\nu}[ \partial^\sigma \partial^\rho h_{\rho\sigma}-(\Box +m^2)h_{\rho}{}^\rho]=0\; . \\ \nonumber\eea

Similarly, the gauge fixed version of the dimensionally reduced  3-form equations, Eqs.~\eqref{BoxA3=10D} can be obtained from the 3-form generalization of the Proca equations
\be\label{3-form=Proca}
\partial^\mu F_{\mu\nu\rho\sigma}+ m^2A_{\nu\rho\sigma}=0\; .
\ee

As far as fermionic fields of the massive counterparts of the type IIA supergravity multiplet are concerned, the gauge fixed form of their equations, Eqs.~\eqref{Dirac=m-10DP}, \eqref{sPsi=0-10D} and  \eqref{divP=0=10D},  can be obtained  from the following massive 10D type IIA counterparts of the Rarita-Schwinger equations

\bea\label{RSm=10D}
\sigma^{\mu\nu\rho}\partial_\nu \Psi_\rho^1 =- m\sigma^{\mu\nu}\Psi_\nu^{2} \; , \qquad \tilde{\sigma}^{\mu\nu\rho}\partial_\nu \Psi_\rho^2 =m\tilde{\sigma}^{\mu\nu} \Psi_{\nu}^{ 1} \; ,
\\ \nonumber \eea
where $\sigma^{\mu\nu\rho}:=\sigma^{[\mu}\tilde{\sigma}{}^{\nu}\sigma^{\rho ]}$, $\tilde{\sigma}{}^{\mu\nu\rho}=\tilde{\sigma}{}^{[\mu}{\sigma}{}^{\nu}\tilde{\sigma}{}^{\rho ]}$,   $\sigma^{\mu\nu}:=\sigma^{[\mu}\tilde{\sigma}{}^{\nu]}$ and  $\tilde{\sigma}{}^{\mu\nu}=\tilde{\sigma}{}^{[\mu}{\sigma}{}^{\nu]}$. In fact, taking the derivative $\partial_\mu$ of these equations, we find
\bea\label{RSm=s2}
\tilde{\sigma}^{\nu\rho}\partial_\nu \Psi_\rho^1 =0\; , \qquad {\sigma}^{\nu\rho}\partial_\nu \Psi_\rho^2 =0
\; .  \
\eea
Then, contracting Eqs.~\eqref{RSm=10D} with $ \tilde{\sigma}^{\mu}$ and $ {\sigma}^{\mu}$, respectively, and using \eqref{RSm=s2} we arrive at
\eqref{sPsi=0-10D}, $\sigma^{\mu}{\Psi}{}_{\mu}{}^{1}=0=
\tilde{\sigma}^{\mu} {\Psi}{}_{\mu }{}^2$. Now Eqs.~\eqref{divP=0=10D}, $\partial^{\mu}
{\Psi}{}_{\mu}{}^{\alpha 1}= 0=
\partial^{\mu}
{\Psi}{}_{\mu \alpha}{}^{ 2}$, follow from \eqref{RSm=s2} and
\eqref{sPsi=0-10D} and, with all these, the RS equation
\eqref{RSm=10D} reduces to the type IIA massive  Dirac equations
\eqref{Dirac=m-10DP},
${\partial}\!\!\!/{}  {\Psi}{}_{\mu}{}^{ 1}=m{\Psi}{}_{\mu }{}^2$ and
 ${\tilde{\partial}}\!\!\!/{}{\Psi}{}_{\mu}{}^2=-m {\Psi}{}_{\mu}{}^{1}$.

Thus, we have shown that the quantum state spectrum of
the D$0$-brane, described by the on-shell superfield \eqref{bDAXi=0} obeying Eqs.~\eqref{D4Xi=ebD4bXi} and \eqref{D4Xi=ebD4bXi}, forms the massive counterpart of type IIA supergravity multiplet. Its linearized equations \eqref{Fierz-Pauli}, \eqref{3-form=Proca} and \eqref{RSm=10D} can be obtained from the following Lagrangian

\bea
\label{L=massiveIIA} \nonumber
L &=& - h^{\mu \nu}
(\Box +m^2) h_{\mu\nu} +2 h^{\mu \nu} \partial^\sigma \partial_{\mu}h_{\nu \sigma}
-h^{\mu \nu}
\partial_{\mu}\partial_{\nu}h_{\rho}{}^\rho  -
h_{\sigma}{}^\sigma \partial^\sigma \partial^\rho h_{\rho\sigma}+h_{\sigma}{}^\sigma(\Box +m^2)h_{\rho}{}^\rho + \\ \nonumber \\
& +&  \frac{1}{4!} F_{\mu \nu \rho \sigma} F^{\mu \nu \rho \sigma}- m^2 \frac{1}{3!}A_{\mu \nu \rho} A^{\mu \nu \rho} +  \nonumber \\ \nonumber \\
&+& i{\Psi}_\mu^1 \sigma^{\mu\nu\rho}\partial_\nu \Psi_\rho^1 +
i {\Psi}_\mu^2 {\tilde \sigma}^{\mu\nu\rho}\partial_\nu \Psi_\rho^2 + 2m i{\Psi}_\mu^1 {\sigma}^{\mu \nu}{ \Psi}_\nu^{2} \\ \nonumber
 \eea
which is invariant under the following supersymmetry transformations
\bea\label{susy=hmn}
\nonumber &&\delta  h_{  \mu \nu } =
-2i{ \Psi}^1_{(\mu}  \sigma_{\nu)} \,\,  \epsilon^1
-2i \bar { \Psi}^2_{(\mu} \tilde \sigma_{\nu)} \,\, \epsilon^2
- \frac {2i} m \partial_{(\mu} \Psi^1_{\nu)}\, \epsilon^2 +  \frac {2i} m \partial_{(\mu} \Psi^2_{\nu)}\, \epsilon^1, \\ \nonumber
\\ \nonumber
&&\delta  A_{  \mu \nu \rho} =
3i {{\Psi}}^1_{[\mu} { \sigma}_{\nu \rho]} \,\, \epsilon^2
-3i
{\Psi}^2_{[\mu}  {\tilde \sigma}_{\nu \rho]} \,\, \epsilon^1 - \frac {6i} m\partial_{[\mu} {{\Psi}}^1_{\nu} { \sigma}_{\rho]} \,\, \epsilon^1
- \frac {6i} m
\partial_{[\mu}{\Psi}^2_{\nu}  {\tilde \sigma}_{\rho]} \,\, \epsilon^2,\\ \nonumber
\\ \nonumber
&&\delta  \Psi_\mu^1 =-2{\tilde \sigma}^{\nu\rho}\,\epsilon^1 \, \partial_\nu h_{\mu \rho}
+2
m {\tilde \sigma}^\nu \epsilon^2 h_{\mu \nu}
- \frac 1 9
\left[\left(\delta_\mu^{[\nu} {\tilde \sigma}^{\rho \tau \lambda]}-\frac 1 8 {\tilde \sigma}_\mu{}^{\nu \rho \tau \lambda} \right)
\, F_{\nu\rho\tau \lambda} - \partial_\mu A_{\rho \tau \lambda} \tilde{\sigma}{}^{\rho \tau \lambda}\right] \,\epsilon^2 +
   \\ \nonumber && {}\qquad - \frac 1 {18}\left(\frac 1 {4m}  \partial_\mu F_{[4]} \tilde{\sigma}{}^{[4]} +  m  A_{[3]} \tilde{\sigma}{}_{\mu}{}^{[3]} - 6mA_{\mu[2]} \tilde{\sigma}{}^{[2]} \right)\, \epsilon^1\; , \\ \nonumber  \\ \nonumber
&&\delta  \Psi_\mu^2 = \; 2\sigma^{\nu\rho}\,\epsilon^2 \, \partial_\nu h_{\mu \rho}
+2 m { \sigma}^\nu \epsilon^1 h_{\mu \nu}
- \frac 1 9
\left[\left(
\delta_\mu^{[\nu} \sigma^{\rho \tau \lambda]}-\frac 1 8 \sigma_\mu{}^{\nu \rho \tau \lambda}
\right)
 F_{\nu\rho\tau \lambda} - \partial_\mu A_{\rho \tau \lambda}  \tilde{\sigma}{}^{\rho \tau \lambda} \right]
\,\epsilon^1\,+
\\  && {}\qquad + \frac 1 {18}\, \left( \frac 1 {4m} \partial_\mu F_{[4]}{\sigma}{}^{[4]}+    m  A_{[3]} {\sigma}{}_{\mu}{}^{[3]} - 6mA_{\mu[2]} {\sigma}{}^{[2]} \right)\epsilon^2 \; . \\ \nonumber\eea

The form of these supersymmetry transformations may look fairly complicated. However, it is straightforward to check that these transformations are consistent with transversality, tracelessness and gamma-tracelesness conditions which follow from equations of motion, as well as with Klein-Gordon and Dirac equations. This implies the invariance of all the equations of motion under this supersymmetry.

\bigskip

\setcounter{equation}0

\section{Conclusions and discussion}
\label{Conclusion}
In this paper, we have developed the Hamiltonian approach to a 10D D$0$-brane in its spinor moving frame formulation,  performed its covariant quantization, identified the quantum state spectrum of  D$0$-brane and found the linearized equations of  D$0$-brane field theory.

The quantization scheme which we have chosen \cite{Bandos:2017zap}, implies the introduction of a complex structure by using auxiliary variables parametrizing $\frac {\text{SO}(9)}{\text{SO}(7)\otimes \text{U}(1)}$ coset (which can be called internal harmonics following the line of \cite{Galperin:1984av,Galperin:1984bu,Galperin:2001seg}) and use the Gupta-Bleuler-type method of quantization for the fermionic constraints. This results in the state vector represented by a chiral (analytic in the terminology of \cite{Galperin:1984av,Galperin:1984bu,Galperin:2001seg}) on-shell superfield. However, as the representation of supersymmetry carried out by such a superfield is reducible, we impose additional duality-type
relation \eqref{D4Xi=ebD4bXi} which, in particular, expresses  the higher component of the chiral on-shell superfield through complex conjugates to its lower components. In this way we make the representation of supersymmetry carried by the worldline superfield irreducible.

Notice that a similar duality-like superfield equation should be imposed on the chiral on-shell superfields describing maximally supersymmetric four dimensional ${\cal N}=8$ supergravity and ${\cal N}=4$ SYM in the frame of the supertwistor approach \cite{Ferber:1977qx}\footnote{These are one-particle counterpart of the superamplitudes of maximally supersymmetric $\text{D}=4$ SYM and ${\cal N}=8$ supergravity widely used in relatively recent break-through of \cite{ArkaniHamed:2008gz,Elvang:2015rqa} and refs. therein.}. We will discuss this and related issues as well as alternative schemes of quantization in the companion paper \cite{in-preparation}.

We have analyzed the quantum state spectrum of the D$0$-brane described by the above-mentioned on-shell superfield and have shown that this describes a massive counterpart of the (linearized) type IIA supergravity\footnote{\label{mIIA} We have tried to avoid calling this ``massive type IIA supergravity'' (although  not rigorously).
The reason is to prevent a possible  confusion with Romans type IIA supergravity \cite{Romans:1985tz} and with Howe-Lambert-West supergravity \cite{Howe:1997qt} which are usually called massive. The quantum state spectrum of D$0$-brane in flat superspace cannot be identified with Romans supergravity as the ground state of this latter is $\text{AdS}_4\times {\mathbb S}^6$ rather than Minkowski spacetime.  Howe-Lambert-West supergravity, which can be obtained from Scherk-Schwarz dimensional reduction of 11D supergravity \cite{Lavrinenko:1997qa} or by topologically nontrivial dimensional reduction of a reformulation of 11D supergravity with a trivial connection for Weyl symmetry \cite{Howe:1997qt}, contains massive vector field in its spectrum.  The unified superfield approach to both these massive supergravities was developed in \cite{Tsimpis:2005vu}, where it is stated that they are only possible massive deformations of type IIA supergravity. An interesting problem for future study is to find a place of ``our'' massive counterpart of type IIA supergravity in the superspace approach of \cite{Tsimpis:2005vu} or its modification.}.
To our best knowledge such a massive counterpart of type IIA supergravity had not been described in the literature, one of the
nontrivial parts of studying the D$0$-brane quantum state spectrum consisted of identification of the field content of
this massive supermultiplet and of its field equations.

Keeping in mind that the standard type IIA supergravity can be obtained by dimensional reduction of 11D supergravity, i.e. appears as leading, massless mode in Kaluza-Klein compactification of 11D supergravity on a circle ${\bb S}^1$, we have found the natural to  expect that its massive counterpart can appear as next-to-leading massive mode in such a decomposition. We have found equations describing such a linearized massive type IIA supergravity,  solved them in terms of spinor frame variables and internal $\text{SO}(9)/\text{SO}(7)$ harmonics, and have established the one-to-one correspondence between parameters of such solutions with components of the on-shell superfield describing the quantum D$0$-brane in its spinor moving frame formulation.

The massive counterpart of  type IIA supergravity multiplet,  which we have found as quantum state spectrum of the D$0$-brane, clearly gives rise to massive graviton and higher spin fields after dimensional reduction to $\text{D}=4$. Thus the field theory of D$0$-brane  belongs to the class of massive higher-spin theories\footnote{A complete list of consistent cubic interaction vertices between massive and massless higher spin fields
is given in \cite{Metsaev:2005ar,Metsaev:2007rn, Metsaev:2012uy} and corresponding on-shell amplitudes were constructed in \cite{Arkani-Hamed:2017jhn}.  A study of properties of interacting massive higher spin fields is an interesting problem in its own right and recently an interesting application has been found. Namely, the scattering amplitudes of massive higher spin fields obtained in \cite{Arkani-Hamed:2017jhn}  were used for an effective field theoretic description of rotating black hole binaries, see for example \cite{Cangemi:2022bew,Skvortsov:2023jbn} and references therein. Notice that theories with massive higher spin fields are different from massless higher spin theories by Vasiliev, also known under the name of higher spin gravity \cite{Vasiliev:1990en,Vasiliev:2003cph,Vasiliev:2025hfh} as well as from chiral higher spin theories, see for example \cite{Ponomarev:2016lrm,Skvortsov:2018jea}. }
 \cite{Bekaert:2022poo}.


Our analysis of the quantum state spectrum D$0$-brane based on its identification with next-to-leading mode of the Kaluza-Klein reduction of 11D supergravity, with mass $m$, together with the fact that the standard  D$0$-brane action can be obtained by similar dimensional reduction of the action for M-wave (which is also called M$0$-brane and is just massless superparticle in  11 dimensional superspace) inspires a question of what is the meaning of the higher Kaluza-Klein modes, with mass $2m$, $3m$, etc.

Notice that when (the ground state of) D$0$-brane with mass $m$ is associated to a supersymmetric solution of type IIA supergravity equations, similar 1/2 BPS solutions  with masses $2m$, $3m$, etc. are usually identified with ground states of multiple D$0$-brane system containing $2$, $3$ etc. D$0$-branes. In contrast, our discussion makes it tempting to think about the  possible relationship of these solutions to higher Kaluza-Klein modes of simplest compactification of 11D M-waves to 10D.
Such a multiplicity of possible associations of supersymmetric point-like and extended objects with a supergravity solution is probably worth further reflection.

The natural application of the results of the present study is to quantize the 10D multiple D$0$-brane (mD$0$)  action from \cite{Bandos:2022uoz} (probably beginning from the warm-up exercise of quantizing the simplest 10D prototype of mD$0$ model first  found in  \cite{Bandos:2018ntt}). This will result in a field theory in the (super)spacetime with additional non-(anti)-commuting coordinates which might provide new insights in String/M-theory.  This hope is related to the special role attributed to the BFSS matrix model \cite{Banks:1996vh} the covariant version of which is provided by mM$0$-action of \cite{Bandos:2012jz} which, in its turn, produces a particular mD$0$ action from \cite{Bandos:2022uoz} in dimensional reduction \cite{Bandos:2022dpx}. The 3D counterpart of this problem was addressed in recently in \cite{Bandos:2023web} by two of the present authors. The study in this paper can be considered as the second step in this direction.

Another interesting direction to develop is related to calculation of scattering amplitudes in type IIA string theory.
For it is important to notice that spinor moving frame formulation of a D$0$-brane provides us with the basic elements  of the spinor helicity formalism for these objects\footnote{The problem of description of scattering of D$0$-branes and strings in the framework of string field theory based on NRS formulation of the superstring was addressed  recently in \cite{Sen:2025xaj}.}. The preliminary study of the 11D (super)amplitude formalism in \cite{Bandos:2017eof,Geyer:2019ayz,Bandos:2019zqp} (see \cite{Caron-Huot:2010nes}  as well as \cite{Dittmaier:1998nn,Conde:2016izb} for relevant earlier studies) suggests to take as fundamental spinor helicity variables the sets of constrained complex spinors

\bea\label{l=mvw}
\bar{\lambda}_{\alpha A}=\sqrt{m} v_{\alpha}{}^{q} \bar{w}_{qA}\; , \qquad  {\lambda}_{\alpha}^A =\sqrt{m} v_{\alpha}{}^{q} w_{q}{}^A\; ,
\eea
composed from the real spinor frame variables \eqref{v=inSpin} and
internal harmonics  \eqref{winSpin=bww}, \eqref{wbw=I}  which we have used above in the D$0$-brane quantization. These obey the constraint

\bea\label{ps=mlbl}
p_{\alpha\beta}=p_\mu\sigma^\mu _{\alpha\beta}= 2{\lambda}_{(\alpha}^A\bar{\lambda}_{\beta)A}\; ,
\eea
which follow from \eqref{ps==mvv}, as well as a number of other constraints. The form of these depends on whether we consider $w$ and $\bar{w}$ variables as $\text{Spin}(9)$ frame (hence indices $A$ transformed by $SO(7)$) or the bridge   between $\text{SO}(9)$ and $\text{SU}(8)$ (in which case the indices $A$ are transformed by $\text{SU}(8)$).

Similarly, we can introduce polarization spinors on the basis of inverse spinor frame matrix \eqref{v-1=inSpin}

\bea
\bar{\lambda}^{\alpha}_{ A}=\sqrt{m} v_{q}{}^{\alpha} \bar{w}_{q A}\; , \qquad  {\lambda}^{\alpha A}=\sqrt{m} v_{q}{}^{\alpha} w_{q}{}^A\; .
\eea
These objects clearly obey

\bea \bar{\lambda}^{\alpha}_{ A}\bar{\lambda}_{\alpha B}=0\; , \qquad \bar{\lambda}^{\alpha}_{ A}{\lambda}_{\alpha}{}^B =\delta_A{}^B \; , \qquad
{\lambda}^{\alpha A}{\lambda}_{\alpha}{}^B  =0\; , \qquad
\delta_\alpha{}^\beta = {\lambda}_{\alpha}{}^A   \bar{\lambda}^{\beta}_{ A}+ \bar{\lambda}_{\alpha A} {\lambda}^{\beta A}\; . \qquad
\eea

Furthermore, using these relations and \eqref{ps=mlbl}, we can easily find the polarization spinors provide us with the basic solutions of the Dirac equations,

\bea
p_{\alpha\beta}\bar{\lambda}^{\beta}_{ A} = m \bar{\lambda}_{\alpha A} \; ,  \qquad p_{\alpha\beta}{\lambda}^{\beta A} = m {\lambda}_{\alpha}^A \; ,  \qquad
\eea
the right hand part of which includes the spinors  helicity variables \eqref{l=mvw}.

On the other hand, we can factorize the timelike momentum in terms of polarization spinors

\bea\label{pts=mlbl}
\tilde{p}{}^{\alpha\beta}=p_\mu\tilde{\sigma}{}^{\mu\alpha\beta}= 2\bar{\lambda}^{(\alpha }_A{\lambda}^{\beta)A}\equiv  2m \bar{v}_A^{(\alpha}{v}^{\beta)A}\;
\eea
and use this relation to present spinor helicity variables as solutions of the Dirac equations,

\be
\tilde{p}{}^{\alpha\beta}\bar{\lambda}_{\alpha A} = m \bar{\lambda}{}^{\alpha }_{A} \, , \qquad \tilde{p}{}^{\alpha\beta}{\lambda}_{\beta}{}^A = m {\lambda}^{\alpha A} \, , \qquad
\ee
so that separation on spinor helicity  variables and polarization spinors seems to be completely conventional in this case.

The form of the above relations, as we have written them, is manifestly  invariant under $\text{SU}(8)$ group acting on the indices $A$, $B$ which were originally treated as $\text{SO}(7)$ spinor indices. As these were introduced in our study as indices of  the internal harmonics $w_{q}{}^{A}$ and $\bar{w}_{qA}$, this observation suggests the possibility to weaken  the constraints which identified these as homogeneous coordinates of $\text{SO}(9)/[\text{SO}(7)\otimes \text{SO}(2)]$ coset and make them a kind of bridge between $\text{SU}(8)$ and $\text{SO}(9)$ group transformations. This suggests a hidden $\text{SU}(8)$ symmetry in our problem.
Actually this hidden symmetry occurs not only in quantum D$0$-brane but more generally in type IIA string theory as well as in 11D supergravity. This issue will be one of the subjects of the companion paper \cite{in-preparation}\footnote{The hidden symmetries of maximal D=4 supergravity theory was observed already in the pioneer paper \cite{Cremmer:1979up} and the problem of its origin in D=11 and D=10 supergravity was addressed in \cite{deWit:1985iy}. In \cite{in-preparation} we will discuss the relation of this hidden $E_7/SU(8)$ symmetry with hidden $SU(8)$ symmetry, which can be observed in linearized 11D, type IIA and type IIB supergravity as well as in linearized D$0$--brane field theory which we described above. Notice that 11D supergravity with local $SU(8)$  constructed in \cite{deWit:1986mz} is different from that in \cite{deWit:1985iy}: it cannot be linearized over the flat 11D superspace and gives rise to gauged ${\cal N}=8$ $D=4$ supergravity upon domensional reduction of ${\bb S}^7$.
}. See Appendix~\ref{sec:SU8inL} for a prequel of this  discussion.

\section*{Acknowledgments}
We would like to thank  Sergei Kuzenko, Yasha Neiman
and especially Dima Sorokin
for useful discussions.
The work of I.B. and U.S. has been supported in part by the MCI, AEI,
FEDER (UE) grants PID2021-125700NB-C21 and PID2024-155685NB-C21  (“Gravity, Supergravity and Superstrings” (GRASS)) and by the Basque Government grant IT-1628-22. The work of M.T. has been supported by the Quantum Gravity Unit of the Okinawa Institute of Science and Technology Graduate University (OIST). I.B. would like to thank the Quantum Gravity Unit of OIST for hospitality during his short visit at an early stage of this work.

\appendix

\section{ SO(7) and SU(8)}
\label{SU8-SO7}
\setcounter{equation}{0}
\def\theequation{\ref{SU8-SO7}.\arabic{equation}}
\subsection{Properties of SO(7) sigma matrices and link to  $SU(8)$ }
Clebsh-Gordan coefficients (generalized Pauli matrices) for $\text{SO}(7)$ are antisymmetric $8\times 8$ matrices with complex coefficients
\be\label{sI=-sIT}
\sigma^{\check{I}}_{AB}= -\sigma^{\check{I}}_{BA}= \sigma^{\check{I}}_{[AB]}\; .
\ee
If we would like to describe the basis of $\text{SU}(8)$ (although not in an explicitly $\text{SU}(8)$ covariant representation), we can consider the set of these matrices and the set of their counterparts with upper indices,
\be\label{ts-chI}
\tilde{\sigma}{}^{\check{I}\; AB}=-\tilde{\sigma}{}^{\check{I}\; BA}\dot{=}~ \tilde{\sigma}{}^{\check{I}\; [AB]}
= - (\sigma^{\check{I}}_{AB})^*=  (\sigma^{\check{I}}_{BA})^*\;
\ee
which obey
\be\label{sts+sts=I}
\sigma^{\check{I}}_{AC}\tilde{\sigma}{}^{\check{J}\; CB} + \sigma^{\check{J}}_{AC}\tilde{\sigma}{}^{\check{I}\; CB}:= 2\sigma^{(\check{I}}_{AC}\tilde{\sigma}{}^{\check{J})\; CB}= \delta_A{}^B\; .
\ee

Then the basis in the space of complex $8\times 8$ matrices with lower indices is provided by $36=35+1$ symmetric

\be\label{s-chI=sym}
 \sigma^{\check{I}_1\check{I}_2\check{I}_3}_{AB}= \sigma^{\check{I}_1\check{I}_2\check{I}_3}_{(AB)}\; , \qquad  \sigma^{\check{I}_1\ldots \check{I}_7}_{AB}= \epsilon^{\check{I}_1\ldots \check{I}_7} {\cal U}_{AB}\; , \qquad
\ee
and $28=7+21$ antisymmetric

\be\label{s-chI=asym}
\sigma^{\check{I}}_{AB}= \sigma^{\check{I}}_{[AB]}\; , \qquad \sigma^{\check{I}_1\ldots \check{I}_5}_{AB}= \sigma^{\check{I}_1\ldots \check{I}_5}_{[AB]}=  \epsilon^{\check{I}_1\ldots \check{I}_5\check{J}_1 \check{J}_2} ({\cal U}\tilde{\sigma}^{\check{J}_1\check{J}_2})_{AB}= \epsilon^{\check{I}_1\ldots \check{I}_5\check{J}_1 \check{J}_2} (\sigma^{\check{J}_1\check{J}_2}{\cal U})_{AB}\; . \qquad
\ee

The complex symmetric matrix $ {\cal U}_{AB}$ defined in  \eqref{s-chI=sym} is unitary, i.e. its conjugate $\bar{{\cal U}}{}^{AB}=({\cal U}_{AB})^*$ is its inverse,
\be\label{cUbcU=1}
  {\cal U}_{AC}\bar{{\cal U}}{}^{CB}=\delta_A{}^B\; . \qquad
\ee

Notice that the symmetric matrices with upper indices are

\be\label{ts-chI=sym}
 \tilde{\sigma}^{\check{I}_1\check{I}_2\check{I}_3\, AB}= \tilde{\sigma}^{\check{I}_1\check{I}_2\check{I}_3\, (AB)}\; , \qquad  \tilde{\sigma}^{\check{I}_1\ldots \check{I}_7\, AB}= \epsilon^{\check{I}_1\ldots \check{I}_7} \bar{{\cal U}}{}^{AB}\;   \qquad
\ee
which is consistent with hermiticity of the generalized Pauli matrices presented in Eq.  \eqref{ts-chI}.

The basis in the space of matrices with two indices of different types is given by

\bea\label{basis=down-up}
\delta_{A}{}^{B}\; , \qquad \sigma^{\check{I}\check{J}}{}_{A}{}^{B}=-  \tilde{\sigma}{}^{\check{I}\check{J}}{}^{B}{}_{A}\; , \qquad  \sigma^{\check{I}_1\ldots \check{I}_4}{}_{A}{}^{B}=  \tilde{\sigma}{}^{\check{I}_1\ldots \check{I}_4}{}^{B}{}_{A} \; , \qquad \\ \nonumber \\
\sigma^{\check{I}_1\ldots \check{I}_6}{}_{A}{}^{B}=  -\tilde{\sigma}{}^{\check{I}_1\ldots \check{I}_6}{}^{B}{}_{A} =\epsilon^{\check{I}_1\ldots \check{I}_6\check{J}}
 \sigma^{\check{J}}_{AC}\bar{{\cal U}}^{CB} =-  \epsilon^{\check{I}_1\ldots \check{I}_6\check{J}}
 {\cal U}_{AC}\tilde{\sigma}^{\check{J}\, CB}  \; . \qquad
\eea
Eqs. \eqref{basis=down-up} imply
\be\label{ts=-UsbU} ({\cal U}\tilde{\sigma}^{\check{J}})_{A}{}^{B}=-  (\sigma^{\check{J}}\bar{{\cal U}})_{A}{}^{B} \qquad \Rightarrow \qquad  \tilde{\sigma}^{\check{J}AB}=   -(\bar{{\cal U}}{\sigma}^{\check{J}}\bar{{\cal U}})^{AB}\; .  \qquad
\ee

Notice that under infinitesimal $\text{SU}(N)$ transformations with parameter $\alpha_B{}^A= (\alpha_A{}^B)^*$
\be\label{vcU=}
\delta {\cal U}_{AB}= 2i \alpha_{(A}{}^C {\cal U}_{B)C}\; , \qquad \delta \bar{{\cal U}}{}^{AB}= -2i
\bar{{\cal U}}{}^{C(A} \alpha_{C}{}^{B)} \; . \qquad
\ee
 The expression for infinitesimal transformations of the matrix ${\cal U}$ which is close to unity is more intuitive 
 
\be
{\cal U}_{AB}= \delta_{AB}+ i u_{AB} \; , \qquad \bar{{\cal U}}{}^{AB}=  \delta^{AB}- i u_{AB}  \; , \qquad u_{AB} = u_{BA}=(u_{AB})^* \;   \qquad
\ee
which reads
\be
\delta u_{AB}= 2\alpha_{(AB)}:= 2 \alpha_{(A}{}^C \delta_{B)C}\; .
\ee

This makes manifest that $\text{SU}(8)$ transformations can be used ``to gauge'' an arbitrary invertible unitary matrix ${\cal U}$ into the unity matrix

\be\label{U=I}
{\cal U}_{AB}~\dot{=}~ \delta_{AB} \; , \qquad \bar{{\cal U}}{}^{AB}~\dot{=}~  \delta^{AB}
\ee
and that this ``gauge'' is preserved by the $\text{SO}(7)$ subgroup of $\text{SU}(8)$. Furthermore, with \eqref{U=I}, Eq.~\eqref{ts=-UsbU} reduces to

\be
\tilde{\sigma}{}^{\check{J}AB}\dot{=} -\sigma^{\check{J}}_{AB}=+\sigma^{\check{J}}_{BA}=(\tilde{\sigma}{}^{\check{J}AB})^\dagger\;
\ee
and lower and upper indices $A,B,\ldots$ become indistinguishable. This latter fact reflects the possibility to identify the $\text{SO}(7)$ charge conjugation matrix with unity matrix.

The above discussion makes manifest that the group which preserves an arbitrary complex symmetric and unitary matrix ${\cal U}_{AB}$ is isomorphic to $\text{SO}(7)$. Such an ${\cal U}_{AB}$ matrix can be considered as (a kind of generalization of the) $\text{SO}(7)$ charge conjugation matrix which is used to relate upper and lower spinor indices.

Notice that any other choice of symmetric unitary  ${\cal U}_{AB}$ can be used to work with $\text{SO}(7)$ vectors and spinor equally well, but keeping the difference between upper and lower $\text{SO}(7)$ spinor indices $A,B,\ldots$ related with the use of matrix  ${\cal U}$ and its c.c. Furthermore, if we do not use the explicit form of the ${\cal U}$-matrix, we can allow for its change and transform upper and lower $A,B,\ldots$  indices by $\text{SU}(8)$ group. This implies a similarity transformations of
$\tilde{\sigma}{}^{\check{J}AB}$ and $\sigma^{\check{J}}_{AB}$
 on their matrix indices, but this does not change their algebra. Hence, in some studies,  we can treat upper and lower  $A,B,\ldots$ as indices of $\text{SU}(8)$ while keeping indices
$\check{I},\check{J}$ etc. of say $\tilde{\sigma}{}^{\check{J}AB}$ as $\text{SO}(7)$ indices. This
way one can see hidden $\text{SU}(8)$ symmetry in
linearized 10D type II and in linearized 11D supergravity.
We postpone a more detailed discussion of this hidden $\text{SU}(8)$ symmetry and type II and 11D linearized supergravity to the forthcoming companion paper \cite{in-preparation}.

It is interesting (and potentially useful) to apply such an approach for generalization of the SO(7) spinor frame formalism proposed in \cite{Bandos:2017zap} for analytic superfield approach to 11D superamplitudes.

\subsection{Representation and some identities for SO(7) Clebsh-Gordan coefficients}
\label{rep-SO7}
A useful representation for 7d generalized Pauli matrices can be given in terms of SO(8) sigma matrices
\be \gamma^i_{A\dot{B}}= \tilde{\gamma}^i_{\dot{B}A}=(\gamma^{\check{I}}_{A\dot{B}}, \gamma^8_{A\dot{B}})\; . \ee (These were denoted by $\gamma^i_{q\dot{p}}$ in \cite{Bandos:2017eof}).
This is given by
\be\label{sSO7=gtg8}
\sigma^{\check{J}}_{AB}= (\gamma^{\check{J}}\tilde{\gamma}^8)_{AB}\; , \qquad
\tilde{\sigma}{}^{\check{J}AB}=  ({\gamma}^8\tilde{\gamma}{}^{\check{J}})_{AB}=
-\sigma^{\check{J}}_{AB}=+\sigma^{\check{J}}_{BA}\; .
\ee

It is also useful to introduce

\bea\label{t44:=SO7}
t^{\check{I}\check{J}\check{K}\check{L}}_{ABCD}&=&
(\sigma^{[\check{I}\check{J}}{\cal U})_{[AB}(\sigma^{\check{K}\check{L}]}{\cal U})_{CD]} \; , \qquad \\ \nonumber \\  \label{tt44:=SO7} \tilde{t}{}_{\check{I}\check{J}\check{K}\check{L}}^{ABCD}&=& (\tilde{\sigma}^{[\check{I}\check{J}|}\bar{{\cal U}})^{[AB}(\tilde{\sigma}^{|\check{K}\check{L}]}\bar{{\cal U}})^{CD]}=(t^{\check{I}\check{J}\check{K}\check{L}}_{ABCD})^*\; , \qquad \\ \nonumber \\ \label{t24:=SO7} t^{\check{I}\check{J}}_{ABCD}&=&
(\sigma^{\check{I}\check{K}}{\cal U})_{[AB}(\sigma^{\check{J}\check{K}}{\cal U})_{CD]} + \sigma^{\check{I}}_{[AB}\sigma^{\check{J}}_{CD]} \; , \qquad  \\ \nonumber \\  \label{tt24:=SO7} \tilde{t}{}_{\check{I}\check{J}}^{ABCD}&=& (\tilde{\sigma}^{\check{I}\check{K}}\bar{{\cal U}})^{[AB}(\tilde{\sigma}^{\check{J}\check{K}]}\bar{{\cal U}})^{CD]}+\tilde{\sigma}^{\check{I}[AB}\tilde{\sigma}^{CD]\check{J}}=(t^{\check{I}\check{J}}_{ABCD})^*\; , \qquad \\ \nonumber
\\ \label{t14:=SO7} t^{\check{I}}_{ABCD}&=&
-(\sigma^{[\check{I}\check{K}}{\cal U})_{[AB}\sigma^{\check{K}]}{}_{CD]}\; , \qquad \\ \nonumber \\  \label{tt14:=SO7} \tilde{t}{}_{\check{I}}^{ABCD}&=&\;  (\tilde{\sigma}^{\check{I}\check{K}}\bar{{\cal U}})^{[AB}\tilde{\sigma}^{}{}^{CD]\, \check{K}}=(t^{\check{I}}_{ABCD})^*\; .
\\ \nonumber
\eea
This set of matrices provides a (pair of) complete basis(es) for the 4-th rank antisymmetric spin-tensors of group(s)
$\text{SO}(7)$ (and $\text{SU}(8)$).

The identities for the $\text{SO}(8)$ matrices which can be found in Appendix A of \cite{Green:1983hw} give rise to the following identity

\bea
&\sigma^{\check{J}}_{AB} \sigma^{\check{J}}_{CD}+{\cal U}_{AB}{\cal U}_{CD} ={\cal U}_{AC}{\cal U}_{BD} - \frac 1 2 \sigma^{\check{J}}_{AC} \sigma^{\check{J}}_{BD}+ \frac 1 4( \sigma^{\check{J}\check{K}}{\cal U})_{AC} ( \sigma^{\check{J}\check{K}}{\cal U})_{BD} \\ \nonumber \\ & \Longrightarrow \qquad  \sigma^{\check{J}}_{A[B} \sigma^{\check{J}}_{CD]} = - \frac 1 2( \sigma^{\check{J}\check{K}}{\cal U})_{A[B} ( \sigma^{\check{J}\check{K}}{\cal U})_{CD]}\; ,
\;  \qquad \\ \nonumber
\eea
as well as to the anti-duality of \eqref{t44:=SO7} and \eqref{tt44:=SO7} and duality of \eqref{t24:=SO7}-\eqref{tt14:=SO7}:

\bea\label{t44=-ebt44}
\tilde{t}{}_{\check{I}\check{J}\check{K}\check{L}}^{ABCD}= -\frac 1 {4!}  \epsilon^{ABCDEFGH}
t^{\check{I}\check{J}\check{K}\check{L}}_{EFGH} \; , \qquad \\ \nonumber  \\ \label{t24=ebt24}
\tilde{t}{}_{\check{I}\check{J}}^{ABCD}= +\frac 1 {4!}  \epsilon^{ABCDEFGH}
t^{\check{I}\check{J}}_{EFGH} \; , \qquad \\ \nonumber\\ \label{t14=ebt14}
\tilde{t}{}_{\check{I}}^{ABCD}= +\frac 1 {4!}  \epsilon^{ABCDEFGH}
t^{\check{I}}_{EFGH} \; . \qquad \\ \nonumber
\eea

Instead of \eqref{t44:=SO7} and \eqref{tt44:=SO7} it is more convenient to use their (anti-)duals with three $\text{SO}(7)$ vector indices

\bea\label{t34:=SO7}
t^{\check{I}\check{J}\check{K}}_{ABCD}&=&
(\sigma^{[\check{I}\check{J}}{\cal U})_{[AB}\sigma^{\check{K}]}{}_{CD]} =  -\frac 1 {4!}  \epsilon^{\check{I}\check{J}\check{K}\check{L}\check{P}\check{Q}\check{R}}
\tilde{t}^{\check{L}\check{P}\check{Q}\check{R}}_{ABCD}
\; , \qquad \\ \nonumber \\  \label{tt34:=SO7} \tilde{t}{}_{\check{I}\check{J}\check{K}}^{ABCD}&=& (\tilde{\sigma}^{[\check{I}\check{J}|}\bar{{\cal U}})^{[AB}\tilde{\sigma}^{CD]\, |\check{K}]}=(t^{\check{I}\check{J}\check{K}}_{ABCD})^*= -\frac 1 {4!}  \epsilon_{\check{I}\check{J}\check{K}\check{L}\check{P}\check{Q}\check{R}}
\tilde{t}_{\check{L}\check{P}\check{Q}\check{R}}^{ABCD}\; , \qquad \\ \nonumber
\eea
which are also anti-dual
\bea\label{t34=-ebt44}
\tilde{t}{}_{\check{I}\check{J}\check{K}}^{ABCD}= -\frac 1 {4!}  \epsilon^{ABCDEFGH}
t^{\check{I}\check{J}\check{K}}_{EFGH} \; . \\ \nonumber
\eea

The identities for $\text{SO}(8)$ gamma (actually sigma) matrices, first presented (to our best knowledge) in \cite{Green:1983hw}, with our representation \eqref{sSO7=gtg8} also imply the duality relation
for contraction of the sigma matrices on their 7-vector indices

\be
\tilde{t}^{ABCD}=\tilde{\sigma}^{\check{I}[AB}\tilde{\sigma}^{CD]\check{I}}= +\frac 1 {4!}  \epsilon^{ABCDEFGH}
\sigma^{\check{J}}_{EF} \sigma^{\check{J}}_{GH}=
+\frac 1 {4!}  \epsilon^{ABCDEFGH}
t_{EFGH}\; .
\ee
We did not include these in the above list of basic elements for decomposition of antisymmetric tensors
because it appears as a trace of symmetric $\text{SO}(7)$ tensor \eqref{t24:=SO7},

\bea
t^{\check{J}\check{J}}_{ABCD}=-t_{ABCD}=- \sigma^{\check{J}}_{[AB} \sigma^{\check{J}}_{CD]} \; , \qquad \tilde{t}_{\check{J}\check{J}}^{ABCD}=-\tilde{t}^{ABCD}=- \tilde{\sigma}^{\check{J}[AB} \tilde{\sigma}^{CD]\check{J}} \; .
\eea

\subsection{SO(7) covariant solution of the linearized equations of massive counterpart of type IIA supergravity}
\label{hIJ-AIJK=SO7}
The above properties  imply that the 4-th rank antisymmetric spin--tensor dual to its c.c.
\bea
\bar{\phi}{}^{ABCD}= +\frac 1 {4!}  \epsilon^{ABCDEFGH}
\phi_{EFGH} =(\phi_{ABCD})^* \; ,
\eea
and hence carrying 84 degrees of freedom (d.o.f.), can be decomposed on {\it real} $\text{SO}(7)$ tensors $(84=35+28+7)$

\bea
i\tilde{t}{}_{\check{I}\check{J}\check{K}}^{ABCD}\phi_{ABCD}\; , \qquad \tilde{t}{}_{\check{I}\check{J}}^{ABCD}\phi_{ABCD}\qquad {\text{and}}\qquad \tilde{t}{}_{\check{I}}^{ABCD}\phi_{ABCD}\; , \qquad
 \; . \qquad
\eea
Further, the symmetric $\text{SO}(7)$ tensor above can be decomposed on its trace and traceless part $(28=1+27)$

\bea
\tilde{t}{}^{ABCD}\phi_{ABCD}= \tilde{\sigma}^{\check{J}[AB} \tilde{\sigma}^{CD]\check{J}}\phi_{ABCD} \qquad \text{and}\qquad \tilde{t}{}_{\check{I}\check{J}}^{ABCD}\phi_{ABCD}+ \frac 1 7 \delta^{\check{I}\check{J}}\tilde{t}{}^{ABCD}\phi_{ABCD}
 \; . \qquad
\eea
(The subtracted term in the last equation is a bit more complex because there is no duality relation between $\tilde{\sigma}^{\check{I}[AB} \tilde{\sigma}^{CD]\check{J}}$ and its. c.c. when $\text{SO}(7)$ indices are not contracted).

All these observations are important for establishing the correspondence between the standard on-shell description of the massive
``graviton'' and massive 3-form of the massive counterpart of type IIA supergravity and the content of the on-shell superfield which we obtain by quantizing the D$0$-brane in its spinor moving frame formulation.

\section{Incorporating the internal spinor frame to the Lagrangian and hidden SU(8) symmetry of quantum D$0$-brane}
\label{sec:SU8inL}
\def\theequation{C.\arabic{equation}}
The aim of  this Appendix is to understand better the issue of degrees of freedom in the $\text{SO}(9)$ spinor frame variables and in its  $\text{SO}(16)$ generalization.
We need
this generalization  in order to introduce the complex structure, used essentially in our quantization procedure. To this end, we would like to consider the possibility to introduce these variables at the level of Lagrangian  mechanics.

Although the main conclusion on the fate of degrees of freedom will be the same for both cases of using of $\text{SO}(9)$ and of $\text{SO}(16)$ harmonics,
we will work with the second case as it also makes
manifest the hidden $\text{SU}(8)$ symmetry of our problem. Actually, as we shall discuss in the companion paper \cite{in-preparation}, this hidden  $\text{SU}(8)$ symmetry is also present in standard linearized type IIA supergravity so that our conclusion on its presence  in massive counterparts of type IIA supergravity suggests  that it can be lifted to type IIA string theory. Similarly, it can be proved that this hidden symmetry is also present in linearized 11D supergravity and presumably can be lifted to M-theory.

Let us firstly introduce the variables
$(w_q^A, \bar{w}_{qA})$ which obey only the constraints \eqref{wbw=I},

\begin{eqnarray}
\label{bww=I}
&& \bar{w}_{qA}\bar{w}_{qB}=0 =w_q^A w_q^B\; , \qquad \bar{w}_{qA} w_q^B=\delta_A{}^B\; , \qquad \bar{w}_{qA} w_p^A +  w_q^A \bar{w}_{pA} = \delta_{qp}  \; . \qquad
\end{eqnarray}
Notice that these conditions imply that $(w_q^A, \bar{w}_{qA})$ variables parametrize the $\text{SO}(16)$ valued matrix
\be\label{w+bw=inSO16}
\left(\begin{matrix}\Re{\rm e}w_q^A \cr \Im{\rm m}w_q^A \cr \end{matrix}\right)  \in \text{SO}(16)\;
\ee
which allows us to call these variables $\text{SO}(16)$ harmonics.

Imposing the conditions \eqref{UIgI=}-\eqref{wgIw=UcU} with constant $\text{SO}(7)$ Pauli matrices \eqref{sI=-sIT}, \eqref{ts=-UsbU} and constant symmetric unitary matrix ${\cal U}$   \eqref{cUbcU=1} we make $(w_q^A, \bar{w}_{qA})$  homogeneous coordinates of $\text{SO}(9)$
(which is possible because $\text{SO}(16)$ has $\text{Spin}(9)$ subgroup) and, taking into account the manifest gauge symmetry of our construction,
homogeneous coordinate of $\text{SO}(9)/[\text{SO}(7)\otimes \text{SO}(2)]$ coset. This allows us to call $(w_q^A, \bar{w}_{qA})$  restricted by  \eqref{wbw=I} and  \eqref{UIgI=}-\eqref{wgIw=UcU} $\text{SO}(9)$ harmonics.

Furthermore, as can be deduced from Appendix~\ref{SU8-SO7},
in same cases, including our description of the quantum state spectrum of the D$0$-brane, we can transform  $w_q^A$ and $\bar{w}_{qA}$ by fundamental and anti-fundamental representations of the $\text{SU}(8)$ group. After that we can maintain (up to rotation from $\text{SO}(7)$ subgroup of $\text{SU}(8)$) all the relations \eqref{UIgI=}-\eqref{wgIw=UcU} defining the $\text{SO}(7)$ vector frame, but with the $\text{SU}(8)$ transformed symmetric unitary matrix  ${\cal U}$. Certainly, then also $\text{SO}(7)$  sigma matrices in these relations are subject to transformations from $\text{SU}(8)/\text{SO}(7)$ on their ``spinor'' indices, but such transformations preserve their algebra including \eqref{sts+sts=I} and \eqref{ts=-UsbU} so that their use is completely legitimate. This is a manifestation of the hidden $\text{SU}(8)$ symmetry of the quantum state spectrum of the D$0$-brane\footnote{As we have already stated above and will show in an accompanying paper \cite{in-preparation}, such a hidden  $\text{SU}(8)$ symmetry is also present in type IIA supergravity and its 11D prototype.}.

The variables obeying \eqref{bww=I} are needed to split the real fermionic coordinate $\Theta_q$ into a pair of two complex fermionic variables conjugated one to another, as in Eq.~\eqref{ThetaA:=}. The real $\Theta_q$ is expressed in terms of these latter by
\be\label{Thq=wThA+}
\Theta^q= \bar{w}_{qA}\Theta^A+\bar{\Theta}_A w_q^A\; . \qquad
\ee

To introduce  the variables  $w_q^A$ and $\bar{w}_{qA}$ constrained by
\eqref{bww=I} at the level of Lagrangian, it is sufficient to
substitute the r.h.s. of \eqref{Thq=wThA+} for $\Theta^q$ into the Lagrangian 1-form \eqref{cLD0=analyt}, thus  arriving at 

\bea\label{cLD0=an-SU8}
{\cal L}_1^{{\rm D}0}=m \left[\text{d}{\rm x}^{{\rm 0}}+i \Theta^A \text{d}\bar{\Theta}_A-i \text{d}\Theta^A \, \bar{\Theta}_A
+ i \Theta^A \left(\bar{w}_{qA}\text{d}\bar{w}_{qB}- \frac 1 4\Omega^{IJ} \bar{w}_{qA}\gamma^{IJ}_{qp}\bar{w}_{pB}
  \right)\Theta^B +
\right. \qquad \nonumber \\ \nonumber \\
\left. + i\bar{\Theta}_A \left({w}_{q}^{A}\text{d}{w}_{q}^{B}- \frac 1 4\Omega^{IJ}{w}^{A}\gamma^{IJ}{w}^{B}
  \right)\bar{\Theta}_B - 2 i \Theta^A \left({w}_{q}^{B}\text{d}\bar{w}_{qA}- \frac 1 4\Omega^{IJ} {w}^B\gamma^{IJ}\bar{w}_{A}
  \right)\bar{\Theta}_B - {\rm x}_{_{An}}^{I} \Omega^{I}\right]  \; ,
\\ \nonumber \eea
and to consider $w_q^A$, $\bar{w}_{qA}$, $\Theta^A$ and $\bar{\Theta}_A$ as independent variables.

To further simplify the Lagrangian form and to clarify its structure, let us notice that the expressions in the brackets multiplied by fermion bilinears contain the $\text{SO}(9)$ covariant derivatives of the $\text{SO}(16)$ harmonics $w_q^A$ and $\bar{w}_{qA}$

\be
\text{d}\bar{w}_{qB}- \frac 1 4\Omega^{IJ}\gamma^{IJ}_{qp}\bar{w}_{pB}  \qquad {\rm and}\qquad \text{d}{w}_{q}^{B}- \frac 1 4\Omega^{IJ}\gamma^{IJ}_{qp}{w}_{p}^{B}\; .
\ee
These can be expressed in terms of (generalized) $\text{SO}(16)$
Cartan forms: complex conjugate

\bea\label{mho=bwdbw+}
\bar{\mho}_{AB}&=& \bar{w}_{qA}\text{d}\bar{w}_{qB}- \frac 1 4\Omega^{IJ}\bar{w}_{qA}\gamma^{IJ}_{qp}\bar{w}_{pB}  \; , \qquad \\ \nonumber \\  \label{bmho=wdw+}
\mho^{AB}&=& {w}_{q}^{A}\text{d}{w}_{q}^{B}- \frac 1 4\Omega^{IJ}{w}_{q}^{A}\gamma^{IJ}_{qp}{w}_{p}^{B}=(\bar{\mho}_{AB})^*\; ,  \qquad \\ \nonumber
\eea
which parametrize $\text{SO}(16)/\text{SU}(8)$ coset and transform covariantly under $\text{SU}(8)$, and imaginary
\bea\label{mho=wdbw+}
{\mho}^A{}_{B}=  {w}_{q}^{A}d\bar{w}_{qB}- \frac 1 4\Omega^{IJ}{w}_{q}^{A}\gamma^{IJ}_{qp}\bar{w}_{pB}  \; ,\qquad
\eea
which transforms as $\text{SU}(8)$ connection. The latter can be used to define the $\text{SU}(8)$ covariant derivative which, for the case of complex fermionic coordinate functions, has the form

\bea
{\cal D}\Theta^A=\text{d}\Theta^A-{\mho}^A{}_{B}\Theta^B \; , \qquad {\cal D}\bar{\Theta}_A=\text{d}\bar{\Theta}_A+\bar{\Theta}_B{\mho}^B{}_{A}\; .
 \eea
In terms of these and covariant Cartan forms from Eqs. \eqref{mho=bwdbw+} and  \eqref{bmho=wdw+} the Lagrangian form \eqref{cLD0=an-SU8} reads

\bea\label{cLD0=an--SU8}
{\cal L}_1^{{\rm D}0}=m \left(d{\rm x}^{{\rm 0}}+i \Theta^A {\cal D}\bar{\Theta}_A-i {\cal D}\Theta^A \, \bar{\Theta}_A
+ i \Theta^A \bar{\mho}_{AB} \Theta^B
+ i\bar{\Theta}_A {\mho}^{AB} \bar{\Theta}_B - {\rm x}_{_{An}}^{I} \Omega^{I} \right) \; . \qquad \\ \nonumber
\eea
which makes manifest its $\text{SU}(8)\otimes \text{SO}(9)$ gauge invariance.

The $\text{SU}(8)\otimes \text{SO}(9)$ covariant derivatives of ${w}_{q}^{A}$ and $\bar{w}_{qB}$  are expressed in terms of the $\text{SO}(16)$ Cartan forms by

\bea
{\cal D}\bar{w}_{qB}&:=& \text{d}\bar{w}_{qB}- \frac 1 4\Omega^{IJ}\gamma^{IJ}_{qp}\bar{w}_{pB} -\bar{w}_{qA}  {\mho}^A{}_{B}= {w}_{q}^{A} \bar{\mho}_{AB} \; , \qquad
\\  {\cal D}{w}_{q}^{A}&:=&\text{d}{w}_{q}^{A}- \frac 1 4\Omega^{IJ}\gamma^{IJ}_{qp}{w}_{q}^{A} + {\mho}^A{}_{B} {w}_{q}^{B}=\bar{w}_{qB}{\mho}^{BA}\; .
\eea
It is straightforward to check that the Cartan forms \eqref{mho=bwdbw+}-\eqref{mho=wdbw+} obey the following Maurer-Cartan equations

\bea
& {\cal D}\bar{\mho}_{AB}  =\text{d}\bar{\mho}_{AB} - \bar{\mho}_{AC} \wedge {\mho}^C{}_{B}+ {\mho}^C{}_{A}\wedge \bar{\mho}_{CB} = - \frac 1 4\frak{F}^{IJ}\bar{w}_{A}\gamma^{IJ}\bar{w}_{B} = \frac 1 4\, \Omega^I\wedge\Omega^J\bar{w}_{A}\gamma^{IJ}\bar{w}_{B}\; , \qquad  \\ \nonumber \\
& {\cal D}{\mho}^{AB}  = \text{d}{\mho}^{AB}  -\mho{}^{A}{}_C \wedge {\mho}^{CB}+ {\mho}^{AC} \wedge {\mho}^{B}{}_C  = - \frac 1 4\frak{F}^{IJ}{w}^{A}\gamma^{IJ}{w}^{B}=  \frac 1 4\, \Omega^I\wedge\Omega^J {w}^{A}\gamma^{IJ}{w}^{B}\; , \qquad \\ \nonumber \\
& \frak{G}^{A}{}_B= \text{d}\mho{}^{A}{}_B +\mho{}^{A}{}_C\wedge \mho{}^{C}{}_B  = {\mho}^{AC}\wedge \bar{\mho}_{CB}-\frac 1 4\, \Omega^I\wedge\Omega^J\bar{w}_{A}\gamma^{IJ}{w}^{B}\; .
\eea

In this way we have used the Maurer-Cartan equations for the $\text{SO}(9)$ group in the form adapted for the description of the $\text{SO}(9)/[\text{SO}(7)\otimes \text{SO}(2)]$ coset,

\bea
\text{D}\Omega^I :=\text{d}\Omega^I +\Omega^J\wedge \Omega^{JI}=0\; , \\ \nonumber \\ \qquad {\frak F}^{IJ}= \text{d}\Omega^{IJ}+\Omega^{IK}\wedge \Omega^{KJ}= - \Omega^{I}\wedge \Omega^{J}\; . \\ \nonumber
\eea
These can be used to obtain (by formal contraction $i_\delta$ with variational symbol $\delta$) the admissible variations of the Cartan forms

\bea
\delta \Omega^I &=&  - \Omega^{IJ} i_\delta \Omega^J +i_\delta \Omega^{IJ} \Omega^J\; , \qquad \\ \nonumber \\
\delta \Omega^{IJ} &=&  -2\Omega^{[I} i_\delta \Omega^{J]}-2\Omega^{[I|K} i_\delta \Omega^{K|J]}\; , \qquad
\\ \nonumber\\ \nonumber
&& \text{and} \\ \nonumber \\  
\delta \bar{\mho}_{AB}& =& 2 \bar{\mho}_{[A|C} i_\delta {\mho}^C{}_{|B]}-2 i_\delta\bar{\mho}_{[A|C} \, {\mho}^C{}_{|B]}+ \frac 1 2\, \Omega^Ii_\delta\Omega^J\bar{w}_{A}\gamma^{IJ}\bar{w}_{B}\; , \qquad \\ \nonumber\\
\delta {\mho}^{AB} & =&  2 {\mho}^{[A|}{}_{C} i_\delta {\mho}^{C|B]}-2 i_\delta {\mho}^{[A|}{}_{C}  {\mho}^{C|B]}+ \frac 1 2\, \Omega^Ii_\delta\Omega^J{w}^{A}\gamma^{IJ}{w}^{B}\; , \qquad \\ \nonumber\\ \delta\mho{}^{A}{}_B &= & i_\delta\mho{}^{A}{}_C \mho{}^{C}{}_B -\mho{}^{A}{}_C i_\delta\mho{}^{C}{}_B  +{\mho}^{AC}i_\delta \bar{\mho}_{CB} - i_\delta{\mho}^{AC} \bar{\mho}_{CB}-\frac 1 2\, \Omega^{[I}i_\delta\Omega^{J]}\bar{w}_{A}\gamma^{IJ}{w}^{B}\; . \\ \nonumber
\eea

The ``admissible'' covariant derivatives with respect  the  $(w_q^A, \bar{w}_{qA})$ restricted by \eqref{bww=I} only can be obtained by decomposing the differential on the $\text{SO}(16)$ group manifold on the basis of covariant $\text{SO}(16)$ and $\text{SO}(9)$ Cartan forms,

\bea
& \text{d}^{(w,\bar{w})}:= \text{d}w_q^A \frac {\partial} {\partial w_q^A} + \text{d}\bar{w}_{qA} \frac {\partial} {\partial \bar{w}_{qA} }= \bar{\mho}_{AB} {\frak D}^{AB}+
{\mho}^{AB}\bar{{\frak D}}_{AB} + {\mho}^{A}{}_{B}{\frak D}^{B}{}_{A}+
\frac 1 2 \Omega^{IJ}{\frak D}^{IJ \, (w,\bar{w})} \; , \qquad
\\ \nonumber \\
& {\frak D}^{AB} ={w}^{[A|}_{q} \frac {\partial} {\partial \bar{w}_{q|B]} }\; ,  \qquad  \bar{{\frak D}}_{AB} = \bar{w}_{[A| q} \frac {\partial} {\partial w_q^{|B]}} \; ,   \qquad  {\frak D}^{B}{}_{A} = \bar{w}_{A q} \frac {\partial} {\partial \bar{w}_{Bq} }-{w}_q^{B}\frac {\partial} {\partial w_q^A} \; , \qquad \\ \nonumber \\
& {\frak D}^{IJ \, (w,\bar{w})}= -\frac 1 2 (\bar{w}_{A }\gamma^{IJ})_q \frac {\partial} {\partial \bar{w}_{qA} }-\frac 1 2 ({w}^{A}\gamma^{IJ})_q \frac {\partial} {\partial w_q^A} \; . \qquad
\\ \nonumber \eea

Notice that these covariant derivatives obey the algebra $\text{so}(16)\oplus \text{so}(9)$ with nonvanishing brackets

\bea \label{fD=alg}
{}[{\frak D}^{AB} , \bar{{\frak D}}_{CD}] &=& \; 2
\delta^{[A}{}_{[C}{\frak D}^{B]}{}_{D]}\; , \qquad
{}[ {\frak D}^{A}{}_{B}, {\frak D}^{CD} ]= \; 2 {\frak D}^{A[C} \delta^{D]}{}_{B} \; , \qquad
{}[ {\frak D}^{A}{}_{B},\bar{{\frak D}}_{CD}]=-2 \bar{{\frak D}}_{B[C} \delta^{A}{}_{D]}
\; , \qquad \\ \nonumber \\ \nonumber && \text{ and } \\ \nonumber \\
{}[{\frak D}^{IJ} , {\frak D}^{KL}] &=&  2
(\delta^{I[K} {\frak D}^{L]J}- \delta^{J[K} {\frak D}^{L]I})
\; .   \\ \nonumber \eea
Similarly, the canonical form in the analytical basis of our enlarged (harmonic) superspace is given by

\bea
& \text{d}{\rm x}^{\rm 0}p_{\rm 0} + \text{d}\Theta^A\Pi_A + \text{d}\bar{\Theta}_A\bar{\Pi}^A
+ \bar{\mho}_{AB} {\frak d}^{AB}+
{\mho}^{AB}\bar{{\frak d}}_{AB} + {\mho}^{A}{}_{B}{\frak d}^{B}{}_{A}-
\frac 1 2 \Omega^{IJ}{\frak d}^{IJ } +
\Omega^{I}{\frak d}^{I } \;  \qquad
\eea
where the covariant momenta are expressed in terms of canonical ones by

\bea
&& {\frak d}^{AB} ={w}^{[A}_{q} P^{\bar{w}}_{q}{}^{B]} \; ,  \qquad  \bar{{\frak d}}_{AB} = \bar{w}_{[A| q} P^{w}{}_{q|B]} \; , \\ \nonumber \\  && {\frak d}^{B}{}_{A} = \bar{w}_{A q} P^{\bar{w} B}_{q} -{w}_q^{B}P^{w}_{qA} \; , \qquad \\ \nonumber\\
&& {\frak d}^{IJ }= {\frak d}_{(v)}^{IJ }+\frac 1 2 (\bar{w}_{A }\gamma^{IJ})_q P^{\bar{w}}_{q}{}^{A}+\frac 1 2 ({w}^{A}\gamma^{IJ})_qP^{w}_{qA}\; , \qquad
 {\frak d}^{I}= {\frak d}_{(v)}^{I } \; , \qquad \\ \nonumber
\eea
and ${\frak d}_{(v)}^{I }$ and ${\frak d}_{(v)}^{I J}$ are covariant momenta for spinor moving frame variables, Eqs. \eqref{fdI:=}.

Calculating these covariant momenta as derivatives of the
Lagrangian 1-form with respect to Cartan forms, 

\bea
 & {\frak d}^{AB}:= \frac 1 2 \, \frac {\partial {\cal L}_1}{\partial \bar{\mho}_{AB}}\; , \qquad
\bar{{\frak d}}_{AB}:= \frac 1 2 \, \frac {\partial {\cal L}_1}{\partial {\mho}^{AB}} \; , \qquad   {\frak d}^{B}{}_{A}:=\, \frac {\partial {\cal L}_1}{\partial {\mho}^{A}{}_{B}}\; , \qquad   {\frak d}^{IJ }:=-  \frac {\partial {\cal L}_1}{\partial  \Omega^{IJ}} \; , \qquad
{\frak d}^{I } :=  \frac {\partial {\cal L}_1}{\partial  \Omega^{I}}  \;  , \\ \nonumber
\eea
and canonical momenta as derivatives of the Lagrangian with respect to velocities,  we find the following set of primary constraints

\bea\label{fdAB==}
{\frak d}^{AB} -im\Theta^A\Theta^B \approx 0 \; , \qquad
\bar{{\frak d}}_{AB} -im\bar{\Theta}_A\bar{\Theta}_B \approx 0  \; , \qquad   {\frak d}^{B}{}_{A}-2im\Theta^B \bar{\Theta}_A \approx 0 \; , \qquad
 \\ \nonumber \\ \label{fdIJ==} {\frak d}^{IJ } \approx 0  \; , \qquad
\\ \nonumber \\ \label{pI-xI}
{\rm x}_{_{An}}^I- \frac 1 m \, {\frak d}^{I }\approx 0 \; , \qquad p_I \approx 0 \; . \qquad
\\ \nonumber \\   \label{p0-m==0}
p_{{\rm 0}} -m\approx 0\; , \qquad
\\ \nonumber \\ \label{fdA=}
\frak{d}_A =\Pi_A+im\bar{\Theta}_A\approx 0\; , \qquad \bar{\frak{d}}{}^A =\bar{\Pi}^A+im\Theta^A\approx 0\; .
\\ \nonumber\eea
Constraint~\eqref{p0-m==0} is clearly of the first class and has vanishing brackets with all other constraints.
 
Two bosonic constraints \eqref{pI-xI} clearly form a pair of resolved second class constraints so that we can simply reduce phase (super)space of our dynamical system \cite{Dirac:1963} by just omitting the corresponding $ 2\times9=18$ directions. Then,  Eq.~\eqref{fdIJ==} becomes the first class constraint which has vanishing brackets with all other constraints. In other words, it generates SO(9) gauge symmetry which leaves all the other constraints inert.

Now, it is convenient to replace the primary constraints
\eqref{fdAB==} by  their linear combinations with the fermionic constraints \eqref{fdA=}:

\bea\label{fdAB=}
\tilde{{\frak d}}{}^{AB}={\frak d}^{AB} + \Theta^{[A}\bar{\Pi}{}^{B]} \approx 0 \; , \qquad
\tilde{\bar{{\frak d}}}_{AB} := \bar{{\frak d}}_{AB} + \bar{\Theta}_{[A}\Pi_{B]} \approx 0  \; , \qquad   \tilde{{\frak d}}{}^{B}{}_{A}=  {\frak d}^{B}{}_{A}+\Theta^B \Pi_A - \bar{\Theta}_A \bar{\Pi}{}^B \approx 0 \; . \qquad
\eea
On the Poisson brackets, the constraints
\eqref{fdAB=}  represent the $\text{so}(16)$ algebra
({\it cf.} \eqref{fD=alg})

\bea
{}[\tilde{{\frak d}}{}^{AB} ,\tilde{\bar{{\frak d}}}_{CD}]_{_{\text{PB}}} &=& - 2
\delta^{[A}{}_{[C}\tilde{{\frak d}}^{B]}{}_{D]}\; , \qquad
{}[ \tilde{{\frak d}}{}^{A}{}_{B}, \tilde{{\frak d}}{}^{CD} ]_{_{\text{PB}}}= - 2 \tilde{{\frak d}}{}^{A[C} \delta^{D]}{}_{B} \; , \qquad
{}[ \tilde{{\frak d}}{}^{A}{}_{B},\tilde{\bar{{\frak d}}}_{CD}]_{_{\text{PB}}}=\;\; 2 \tilde{\bar{{\frak d}}}_{B[C} \delta^{A}{}_{D]}
\; \\ \nonumber \eea
and generate SO(16) gauge symmetry.  This transforms the pair of fermionic constraints \eqref{fdA=} covariantly as show

\bea
{}[\tilde{\bar{{\frak d}}}_{AB},{\frak d}_{C}]_{_{\text{PB}}} =0\; , \qquad
{}[\tilde{{\frak d}}{}^{AB} ,{\frak d}_{C}]_{_{\text{PB}}} = \bar{{\frak d}}{}^{[A}
\delta^{B]}{}_{C}\; , \qquad {}[ \tilde{{\frak d}}{}^{A}{}_{B},   ,{\frak d}_{C}]_{_{\text{PB}}} =  {\frak d}_{B} \delta^{A}{}_{C}
\;  \qquad \\ \nonumber 
\eea
and their c.c. relations. The remaining fermionic constraints \eqref{fdA=} are 8 pairs of canonically conjugate and (minus) complex conjugate second class constraints. We can deal with these as in the main text, using the generalized Gupta-Bleuler (or conversion) method.

The important conclusion from the above discussion is that,
after incorporating SO(16) harmonic variables $(w_q^A, \bar{w}_{qA})$ of Eq.~\eqref{w+bw=inSO16}  into the Lagrangian of the D$0$-brane in the analytical basis, Eq.~\eqref{cLD0=analyt}, we arrive at the Lagrangian form \eqref{cLD0=an-SU8} (equivalent to \eqref{cLD0=an--SU8}) which possesses SO(16) gauge symmetry as well as SO(9) gauge symmetry. The SO(16) makes newly introduced variables $(w_q^A, \bar{w}_{qA})$ pure gauge while  the preserved SO(9) gauge symmetry is important as an identification relation for the spinor moving frame variables.

This  proves the consistency of the approach used in the main text where the introduction of new dynamical variables
$(w_q^A, \bar{w}_{qA})$ (their restricted to be SO(9) harmonics) were used for the quantization procedure, without discussing whether they carry (or bring) new  degrees of freedom.

Of course, we can use the $\text{SO}(16)$ (or second $\text{SO}(9)$) gauge symmetry to fix $(w_q^A, \bar{w}_{qA})$ at some fixed constant values, e.g.
$\Re{\rm e}(w_q^A)=  \delta_{q}^A$, $\Im{\rm m}(w_q^A)=  \delta_{q}^{A+8}$ (or ``square roots'' of constant $\text{SO}(9)$ vectors $\bar{U}_I= (\delta_I^8+i\delta_I^9)=(U_I)^*$ in the sense of \eqref{UIgI=}), and use these constant blocks to perform quantization. However, after such  gauge fixing, the transformations of $\text{SO}(9)$ gauge symmetry which was used originally as identification relation on the set of spinor moving frame variables, mix $w_q^A$ and $\bar{w}_{qA})$ in its $\text{SO}(9)/\text{SO}(7)$ part so that the invariance of the quantization procedure is not apparent.

In fact, what makes this invariance manifest is the above approach in which the additional dynamical variables $(w_q^A, \bar{w}_{qA})$, parametrizing either the
$\text{SO}(16)$ or
the $\text{SO}(9)$ group, are involved in the Lagrangian.

\bigskip

\end{widetext}

\end{document}